\documentclass[aps,prb,twocolumn,amsmath,amssymb,footinbib,showpacs,longbibliography,superscriptaddress,10pt]{revtex4-1}

\bibpunct{[}{]}{;}{n}{}{}

\usepackage[english]{babel}
\usepackage{latexsym}
\usepackage{graphics}
\usepackage{subfigure}
\usepackage{epsfig}
\usepackage{xcolor}
\usepackage{hyperref}
\usepackage{braket} 
\usepackage[T1]{fontenc}
\usepackage[latin9]{inputenc}
\usepackage{amstext}
\usepackage{amssymb}
\usepackage{graphicx}
\usepackage{esint}
\usepackage{braket}
\usepackage{babel}
\usepackage{amsmath}
\usepackage{comment}
\hypersetup{
colorlinks=true,
citecolor=blue,
linkcolor=red,
urlcolor=black
}


\begin{document}

\title{Breakdown of Linear Spin-Wave Theory in a Non-Hermitian Quantum Spin Chain}

\author{Julien Despres}
\affiliation{Universit\'e Paris-Saclay, CNRS, LPTMS, 91405, Orsay, France}

\author{Leonardo Mazza}
\affiliation{Universit\'e Paris-Saclay, CNRS, LPTMS, 91405, Orsay, France}

\author{Marco Schir\`o}
\affiliation{JEIP, USR 3573 CNRS, Coll\`ege de France, PSL Research University, 11 Place Marcelin Berthelot, 75321 Paris Cedex 05, France}

\date{\today}

\begin{abstract}
We present the spin-wave theory of the excitation spectrum and quench dynamics of the non-Hermitian transverse-field Ising model.
The complex excitation spectrum is obtained for a generic hypercubic lattice using the linear approximation of the Holstein-Primakoff transformation together with the complex bosonic Bogolyubov transformation. 
In the one-dimensional case, our result compares very well 
with the exact quasiparticle dispersion relation obtained via a fermionic representation of the problem, at least in the regime of large dissipation and transverse field. When applied to the quench dynamics we show however that the linear spin-wave approximation breaks down and the bosonic theory is plagued by a divergence at finite times. We understand the origin of this instability using a single mode approximation. While limited to short times, we show that this approach allows us to characterize the dynamics arising from the quench of the dissipative term and the light-cone structure of the propagation quantum correlations. Furthermore, for the one-dimensional case, the linear spin-wave dynamics shows good agreement with the exact fermionic solution, both for  the local magnetization and the spin-spin correlations. 
\end{abstract}

\maketitle

\section{Introduction}
\label{sec:intro}

In the last decades, the significant and major advances in the experimental control of quantum matter have given dramatic impulse to the experimental and theoretical investigation
of the out-of-equilibrium dynamics of isolated quantum lattice models  \cite{jurcevic2014, richerme2014, cheneau2012, polkovnikov2011, gogolin2016, calabrese2005}. However, since a quantum system cannot be perfectly isolated from its environment, the study of open quantum systems has attracted a lot of interest in the last few years by taking into account dissipative effects such as local dephasing noise 
\cite{Eisler_2011,kollath2018,turkeshi2021diffusion}, local gain/loss processes \cite{alba2022, rosso2021, rosso2023,mazza2023dissipative} and incoherent hopping \cite{alba2020}. These quantum lattice models are governed by the Lindblad master equation \cite{breuer2007} describing the time evolution of the density matrix associated to the quantum system. The latter contains a unitary part representing the standard 
time evolution of the quantum system with its corresponding Hermitian Hamiltonian, but also a non-unitary part describing the dissipation which is represented by the so-called jump operators. In the no-click limit i.e.~ when no quantum jump occurs, the Lindblad master
equation can be rewritten as the time evolution of a quantum system governed by a non-Hermitian Hamiltonian \cite{ashida2020}. Its non-Hermiticity is responsible for the non-unitary time evolution of the open many-body quantum system \cite{daley2014}. \\

\indent Non-Hermitian quantum many-body systems host unexpected novel and rich physical properties compared to their Hermitian counterparts. Their spectral features have been the focus of several studies related to the existence of exceptional points~\cite{Heiss_2012} and their associated topological properties~\cite{gong2018topological,bergholtz2021exceptional, chen2023}. More recently their entanglement structure  has drawn attention, both for what concerns eigenstates in the biorthogonal formulation~\cite{couvreur2017entanglement,herviou2018entanglement,chang2020entanglement} and for dynamics, where several entanglement transitions have been reported~\cite{sarang2021,turkeshi2023,legal2023,kawabata2023entanglement,granet2023volume,su2023dynamics,zhang2023antiunitary}. Similarly, the non-unitary dynamics after a quantum quench or a Floquet driving has also been studied~\cite{dora2020quantum,banerjee2023emergent}. While most of the literature has focused on many-body \emph{non-interacting systems}, less is known about the behavior of interacting non-Hermitian quantum many body systems. One dimensional models, such as non-Hermitian spin chains or the interacting version of Hatano-Nelson model~\cite{hatano1996localization}, have been studied using  exact diagonalization~\cite{zhang2022symmetry,starkov2023quantum,ghosh2023hilbert} and extensions of Bethe Ansatz and bosonization~\cite{yamamoto2022universal,kattel2023exact,durr2009}.

In this respect it is important to develop theoretical methods for non-Hermitian systems, which are able to capture their spectral and dynamical properties in particular in dimensions higher than one. For quantum spin models a textbook approach is offered by spin-wave theory within the Holstein-Primakoff transformation~\cite{holstein1940field,dyson1956general}, that provides a way to characterize the quasiparticle spectrum of various interacting systems in the equilibrium low-energy limit and which has been successfully extended to the unitary dynamics~\cite{sandri2012linear,lerose2019impact}. The goal of this work is to develop a spin-wave theory for a prototypical non-Hermitian quantum spin chain, the Ising model in a complex transverse field, in one, two and three dimensional lattices. We find that while a straightforward generalization of linear spin-wave theory captures well the spectrum of non-Hermitian quasiparticles, at least for strong values of dissipation and transverse field, it fails dramatically in describing the non-unitary dynamics at long times, particularly in cases where the initial state and the long-time steady state are not smoothly connected by a small number of quasiparticle excitations (spin flips). While one might have expected in this case the non-Hermitian dynamics to lead to a growth (amplification) of quasiparticle occupation numbers, thus quickly leaving the regime of validity of linear spin wave theory, we show in fact that the breakdown of linear spin-wave theory is much more severe and leads to a finite-time divergence of the dynamics. We provide a simple analytical understanding of this instability in terms of a single mode bosonic model. Although limited to short-times, we show that our spin-wave analysis allows us to discuss the dynamics of correlation spreading in the quantum Ising chain. This topic has been extensively investigated for isolated quantum lattice models \cite{despres2018, despres2019, despres2021, barmettler2012, hauke2013, natu2013, cevolani2015, cevolani2016, buyskikh2016, frerot2018}, but is still debated in the specific case of non-Hermitian quantum lattice models and a large number of points remain unanswered \cite{meden2023, ashida2018, moca2021}. For one dimension we benchmark this approach with the exact solution obtained via Jordan-Wigner transformation into non-interacting non-Hermitian fermions~\cite{lee2014bis,turkeshi2023}, confirming the validity of the proposed approach for the spectrum and for the short-time dynamics. \\

This article is organized as follows: in Sec.~\ref{sec:model} we start by introducing the non-Hermitian Ising model and we apply the linear spin-wave theory to deduce its excitation spectrum. Then, in Sec.~\ref{sec:dynamics} we investigate the quench dynamics of the resulting  bosonic non-Hermitian model using the equations of motion for the single particle correlation matrix. We highlight a dynamical instability at finite time in the bosonic theory that we understand using a single mode model. Using this theory we discuss the short-time spreading of spin-spin correlation functions and the local magnetization in one and two dimensional lattices, and compare with the exact fermionic solution in the former case. Finally in Sec.~\ref{sec:conclusion} we present our conclusions. In the Appendices we include further technical details on our results.

%

\section{Model and spin-wave theory for the excitation spectrum}
\label{sec:model}

\subsection{One-dimensional lattice}
We focus on the spin-$1/2$ non-Hermitian transverse-field Ising model (TFIM) on a one-dimensional chain of $N_s$ sites whose lattice spacing is 
fixed to unity ($a=1$); for simplicity, we also set $\hbar = 1$. 
The Hamiltonian $\hat{H}$ reads:
\begin{equation}
\hat{H} =  J \sum_R \hat{S}^x_R \hat{S}^x_{R+1} - (h + i\gamma) \sum_R \hat{S}^z_R,
\label{H_ising_chain}
\end{equation}
where $\hat{S}^{\alpha}_R$ refers to the spin operator acting on the lattice site $R$ along the $\alpha \in \{x,z\}$ axis. 
$J > 0$ denotes 
the antiferromagnetic coupling between nearest neighbours spins, $h > 0$ is the constant and homogeneous transverse magnetic field and $\gamma $ is the dissipative strength representing the local dephasing noise. From the perspective of open quantum systems described by a Lindblad master equation~\cite{breuer2007}
the latter can be interpreted as a heating process arising from a jump operator $\hat{L}_R = \sqrt{2\gamma}(1/2 + \hat{S}^z_R)$ when $\gamma>0$, leading to a non-Hermitian Hamiltonian of the form in Eq.~(\ref{H_ising_chain}), see Appendix ~\ref{lindblad_appendix} for further details on the derivation of the non-Hermitian Hamiltonian in the so called no-click limit.
 In most of the paper we will take $\gamma > 0$, but in Sec.~\ref{Sec:QuenchNegativeGamma} we will present a set of data obtained for $\gamma <0$, which corresponds to a different dissipative process with jump $\hat{L}_R = \sqrt{2|\gamma|} (1/2- \hat{S}^z_R)$.
 In what follows, we will mainly discuss the properties of the paramagnetic phase with spins aligned with the magnetic field for $h \gg J$. In most of our study we will also take $|\gamma| \gg J$.
For $J=0$ the properties of this non-Hermitian Hamiltonian are easy to discuss because the sites are decoupled. Let us focus on a single site for simplicity. The $\ket{\uparrow}$ and $\ket{\downarrow}$ states are eigenvectors of $\hat H$ with eigenvalues $\mp \frac{1}{2} (h+i \gamma)$. If we take $\ket{\uparrow}$ as a reference state, we can thus introduce a complex excitation spectrum equal to $(h + i \gamma)$. The goal of this section is to discuss how this is modified by the presence of the antiferromagnetic coupling $J$.

In order to deduce the excitation spectrum of the non-Hermitian TFIM, we rely on the Holstein-Primakoff transformation:
\begin{subequations}
\begin{align}
& \hat{S}^+_R = \left(1 - \hat{a}^{\dag}_R \hat{a}_R\right)^{1/2} \hat{a}_R, \label{hpp}\\
& \hat{S}^-_R = \hat{a}^{\dag}_R \left(1 - \hat{a}^{\dag}_R \hat{a}_R\right)^{1/2}, \label{hpm} \\
& \hat{S}^z_R = 1/2 - \hat{a}^{\dag}_R \hat{a}_R; \label{hpz}
\end{align}
\end{subequations}
where the $\hat{a}_R$ ($\hat{a}^{\dag}_R$) are annihilation (creation) bosonic operators at the lattice site $R$ that
obey canonical commutation relations. While this is an exact mapping between spin $1/2$ and bosons, the presence of the square root makes the transformation non-linear and of limited practical use since the Hamiltonian of the TFIM written in the bosonic language is highly non-linear. However, in the regime in which the bosonic occupation number remains small, $\langle \hat{a}^{\dag}_R \hat{a}_R \rangle \ll 1$, which is the case for example in the paramagnetic phase of the TFIM in equilibrium, one can expand the square root entering the Holstein-Primakoff transformation and write to leading order $\hat{S}^+_R = \hat{a}_R$ and $\hat{S}^-_R = \hat{a}^{\dag}_R$ so that
\begin{equation}
\hat{S}^x_R = \frac{\hat{a}_R + \hat{a}^{\dag}_R}{2},~~~\hat{S}^{y}_R = \frac{\hat{a}_R - \hat{a}^{\dag}_R}{2i},~~~\hat{S}^z_R = \frac{1}{2}-
\hat{a}^{\dag}_R \hat{a}_R. 
\end{equation}
This is the essence of the linear spin wave theory, which leads to a quadratic bosonic Hamiltonian that is best expressed in momentum space (see Appendix \ref{fourier_transform_op}):
\begin{equation}
\hat{H} = \frac{1}{2} \sum_k A_k \left(\hat{a}^{\dag}_k \hat{a}_k + \hat{a}_{-k}\hat{a}^{\dag}_{-k} \right) + 
B_k \left(\hat{a}^{\dag}_k \hat{a}^{\dag}_{-k} + \hat{a}_k \hat{a}_{-k} \right), 
\label{H}
\end{equation}
with $A_k = h + (J/2)\cos(k) + i\gamma$ and $B_k = (J/2) \cos(k)$. The momentum-dependent complex function $A_k$ is responsible for
the non-hermiticity of the Hamiltonian. 
Hamiltonian~\eqref{H} is the spin-wave theory of the non-Hermitian TFIM.
\begin{figure}[t]
\hspace{-0.5cm}
\includegraphics[scale = 0.44]{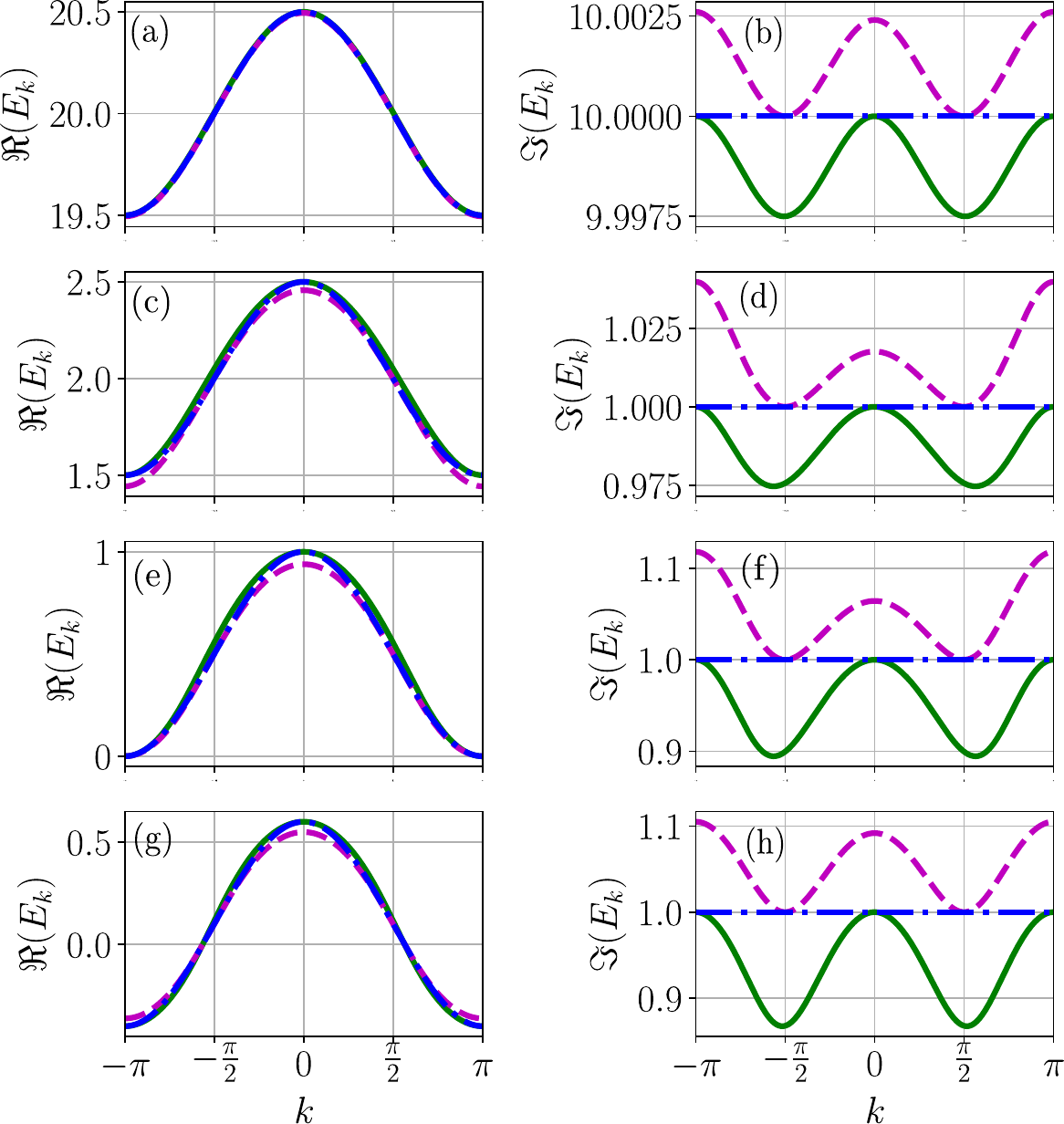}
\caption{Analytical results concerning the complex excitation spectrum $E_k$ of the non-Hermitian transverse Ising chain in the $z$ polarized phase using the bosonic approach relying on the linear approximation to the Holstein-Primakoff transformation, together with the complex bosonic Bogolyubov transformation. The real and imaginary part of the excitation spectrum $E_k$ denoted by $\Re(E_k)$ and $\Im(E_k)$, deduced from Eq.~\eqref{spectrum}, are represented as dashed magenta lines on the panels (a,c,e,g) and (b,d,f,h) respectively. The excitation spectrum $E_k$ is compared to the momentum-dependent function $A_k$ given by $A_k = h + (J/2)\cos(k) + i\gamma$, see dashed-dotted blue lines on panels (a,c,e,g) and (b,d,f,h) for the real and imaginary part of $A_k$ respectively. The complex quasiparticle dispersion relation $E_k$ is also compared to the complex excitation spectrum $\epsilon_k$ found using the fermionic approach relying on the Jordan-Wigner transformation together with the complex fermionic Bogolyubov theory. The real and imaginary part of the excitation spectrum $\epsilon_k$, deduced from Eq.~\eqref{spectrum_2}, are represented as solid green lines on the panels (a,c,e,g) and (b,d,f,h) respectively. The different set of parameters are $J=1$, $h=20$, $\gamma = 10$ for the panels (a,b), $J=1$, $h=2$, $\gamma=1$ for the panels (c,d), $J=1$, $h=0.5$, $\gamma=1$ for the panels (e,f) and $J=1$, $h = 0.1$, $\gamma = 1$ for the panels (g,h).}
\label{plot1}
\end{figure}

A complex bosonic Bogolyubov transformation \cite{lee2014} diagonalizes the non-Hermitian Hamiltonian at Eq.~\eqref{H}, which reads:
\begin{subequations}
\begin{align}
\hat{a}_k^{\dag} =& \bar{\hat{b}}_k \cosh\left( \frac{\theta_k}{2} \right) + \hat{b}_{-k}\sinh\left( \frac{\theta_k}{2} \right), \label{b1}\\
\hat{a}_k =& \bar{\hat{b}}_{-k}\sinh\left( \frac{\theta_k}{2} \right) + \hat{b}_k \cosh\left( \frac{\theta_k}{2} \right), \label{b2}
\end{align}
\end{subequations}
where $\theta_k \in \mathbb{C}$ is an even function of $k$ and $\hat{b}_k$ ($\bar{\hat{b}}_k$) are annihilation (creation) bosonic operators for a quasimomentum $k$ satisfying:
\begin{equation}
 [\hat{b}_k, \bar{\hat{b}}_{k'}] = \delta_{k,k'}, \quad 
 [\hat{b}_k, \hat{b}_{k'}] = [\bar{\hat{b}}_k, \bar{\hat{b}}_{k'}] = 0.
\end{equation}
They define a vacuum state $\ket{0}$ via $\hat{b}_k \ket{0} = 0$ that, in the Hermitian case, corresponds to the ground state; here, it is rather a reference state for the bosonic expansion.
Note that $\bar{\hat{b}}_k \neq \hat{b}_k^{\dag}$ because $\theta_k= \mathrm{arctanh}\left(-B_k/A_k\right) \in \mathbb{C}$. 
The diagonal form of $\hat{H}$ reads:
\begin{equation}
\hat{H} = \frac{1}{2}\sum_k E_k \left( \bar{\hat{b}}_k \hat{b}_k + \hat{b}_{-k}\bar{\hat{b}}_{-k}\right),
\label{H_diag}
\end{equation} 
where the complex excitation spectrum $E_k$ is given by
\begin{equation}
 E_k = \mathrm{sgn}(\Re(A_k)) \sqrt{A_k^2-B_k^2},
\label{spectrum} 
\end{equation}
and is plotted on Fig.~\ref{plot1}.

When $J \ll h, \gamma$, the limit of our interest, it is possible to perform a Taylor expansion of the spectrum:

\begin{equation}
E_k \approx (h+ i \gamma)+ \frac{J}{2}\cos(k) - \frac{J^2}{8(h+ i \gamma)} \cos^2 (k).
\end{equation}

Fig.~\ref{plot1} highlights the fact that in general even an approximation at first order in $J$ gives a good qualitative agreement for the real part of the excitation spectrum, $\Re (E_k) \approx h + \frac{J}{2} \cos (k)$. For $h > J/2$ the spectrum is gapped whereas for $h \leq J/2$ it becomes gapless. 
In order to describe the momentum dependence of the imaginary part of the excitation spectrum, it is instead necessary to go at second order in $J$, since the first order correction is real. We obtain $\Im (E_k) \approx \gamma \left(1 + \frac{J^2}{8 (h^2+ \gamma^2)}\cos^2(k) \right)$ and we observe that we obtain no stable modes in the limit $h,\gamma \gg J$.

Remarkably, the first order approximation yields a good approximation of the data even when $J$ is of the order of $h$ and $\gamma$.
Even if it is tempting to use the bosonic model in this regime of parameters, one should always remember that the approximation based on the linearization of the Holstein-Primakoff transformation is valid only for $h, \gamma \gg J$.


\subsection{One-dimensional lattice: Comparison with the exact fermionic approach.}

In order to assess the validity of the spin-wave theory developed in the previous paragraph, we benchmark it with the exact results that can be obtained in the one-dimensional case via the Jordan-Wigner transformation and the mapping to fermions.
In this case, the excitation spectrum of the non-Hermitian TFIM can be obtained via a complex fermionic Bogolyubov theory.



We introduce the fermionic creation (annihilation) operators $\hat{c}^{\dag}_R$ $\left( \hat{c}_R \right)$  for the lattice site $R$ that satisfy the canonical anti-commutation relations. 
They can be related to spin-1/2 operators via the Jordan-Wigner transformation:
\begin{subequations}
\begin{align}
& \hat{S}^x_R = \frac{1}{2} \hat{K}_R \left( \hat{c}^{\dag}_R + \hat{c}_R \right), \\
& \hat{S}^y_R = \frac{i}{2} \hat{K}_R \left( \hat{c}^{\dag}_R - \hat{c}_R \right), \\
& \hat{S}^z_R = \frac{1}{2} - \hat{c}^{\dag}_R \hat{c}_R,
\end{align}
\end{subequations}
with the non-local string operator $\hat{K}_R = \prod_{R'=0}^{R-1} \left(1 - 2\hat{c}^{\dag}_{R'}\hat{c}_{R'} \right)$.

In this novel language, the one-dimensional non-Hermitian TFIM in Eq.~\eqref{H_ising_chain} is a quadratic function of the fermionic operators (here in momentum representation, see Appendix
\ref{fourier_transform_op}):
\begin{equation}
\hat{H} = \sum_k \mathcal{A}_k \left( \hat{c}^{\dag}_k \hat{c}_k - \hat{c}_{-k} \hat{c}^{\dag}_{-k} \right) + 
\mathcal{B}_k \left( \hat{c}^{\dag}_k \hat{c}^{\dag}_{-k} + \hat{c}_k \hat{c}_{-k} \right),
\label{H_fermionic_quadratic}
\end{equation}

\noindent
where the momentum-dependent functions $\mathcal{A}_k$ and $\mathcal{B}_k$, responsible for the non-hermiticity of $\hat{H}$, are defined as follows  \cite{hickey2013, lee2014bis, turkeshi2023}:
$\mathcal{A}_k = (J/4)\cos(k) + (h+i\gamma)/2$ and $\mathcal{B}_k = -i(J/4)\sin(k)$. We rely on a complex fermionic Bogolyubov transformation \cite{zhang2012} to diagonalize 
the quadratic Hamiltonian $\hat{H}$ at Eq.~\eqref{H_fermionic_quadratic} defined by 
\begin{align}
& \hat{c}^{\dag}_{-k} = \cos \left( \frac{\theta_k}{2} \right) \bar{\hat{\eta}}_{-k} + i \sin \left( \frac{\theta_k}{2} \right) \hat{\eta}_k, \label{tr_1}\\
& \hat{c}_k = \cos \left( \frac{\theta_k}{2} \right) \hat{\eta}_k + i \sin \left( \frac{\theta_k}{2} \right) \bar{\hat{\eta}}_{-k}, \label{tr_2}
\end{align}
where $\theta_k \in \mathbb{C}$ is an odd function of $k$. $\hat{\eta}_k$ $\left( \bar{\hat{\eta}}_k \right)$ denotes the annihilation (creation) fermionic Bogolyubov operator for a quasimomentum $k$.
They fulfill the canonical anti-commutation rules for fermions given by $\{ \hat{\eta}_k, \bar{\hat{\eta}}_{k'} \} = \delta_{k,k'}$, $\{ \hat{\eta}_k, \hat{\eta}_{k'} \} = \{ \bar{\hat{\eta}}_k,
\bar{\hat{\eta}}_{k'} \} = 0$. Besides, they define the vacuum state $\hat{\eta}_k \ket{0} = 0$. It is crucial to underline
that $\bar{\hat{\eta}}_k \neq \hat{\eta}^{\dag}_k$ because $\theta_k$ is complex.

After applying the complex fermionic Bogolyubov transformation, we find
$\theta_k = \mathrm{arctan}\left(i \mathcal{B}_k / \mathcal{A}_k \right)$, see a plot in Appendix~\ref{theta_k_appendix} where we make a comparison with the bosonic case, and the Hamiltonian $\hat{H}$ becomes diagonal in terms of $\hat{\eta}_k$,
$\bar{\hat{\eta}}_k$ and has the following expression
\begin{equation}
\hat{H} = \sum_k \epsilon_k \bar{\hat{\eta}}_k \hat{\eta}_k, 
\end{equation}
where the complex excitation spectrum $\epsilon_k$ has the following expression 
\begin{equation}
\epsilon_k = 2 \mathrm{sgn}(\Re(\mathcal{A}_k))\sqrt{\mathcal{A}_k^2 - \mathcal{B}_k^2}.
\label{spectrum_2}
\end{equation}

In the paramagnetic phase $J \ll h, \gamma$, it can be expanded as:

\begin{equation}
 \epsilon_k \approx  (h + i \gamma) + \frac{J}{2} \cos(k) + \frac{J^2}{8(h + i \gamma)} \sin^2 (k).
\end{equation}

\noindent
Hence, we get $\Im(\epsilon_k) \approx \gamma \left(1 - \frac{J^2}{8(h^2 + \gamma^2)}\sin^2(k) \right)$ and we obtain no stable modes in the limit $h,\gamma \gg J$. 
We observe that the results deduced from the bosonic spin wave theory agree at first order in $J$ with the exact fermionic spectrum. This means that at the level of the real part of the excitation spectrum the two theories are basically identical.
It is interesting to observe the qualitative resemblance of the imaginary part of the spectrum in the fermionic and bosonic cases. The momentum dependence, captured in both cases by the term proportional to $J^2$, in one case is proportional to $+ \cos^2 (k)$ and in the other to $- \sin^2 (k)$. Although this leads to a quantitative disagreement, the qualitative momentum dependence is the same.
Fig.~\ref{plot1} presents a quantitative comparison of the bosonic spectrum with the exact fermionic one.

\subsection{Generalization to a square and hypercubic lattice} \label{Sec:Exc:Sp:2D}

Having benchmarked the spin-wave theory for the non-Hermitian TFIM with the exact result, we can now generalised it to higher-dimensional lattices, where exact approaches are not known.
We begin for simplicity with the case of a square lattice:
\begin{equation}
\hat{H} = J \sum_{\bold{R}}\left( \hat{S}^x_{\bold{R}} \hat{S}^x_{\bold{R} + \bold{x}} + \hat{S}^x_{\bold{R}} \hat{S}^x_{\bold{R} + \bold{y}} \right)
- (h+i\gamma)\sum_{\bold{R}}\hat{S}^z_{\bold{R}},
\label{ham_2D}
\end{equation}
with $\bold{x} = (a_x ~~0)^T$ and $\bold{y} = (0 ~~a_y)^T$ and where $a_x = a_y = a = 1$. 

\begin{figure}[t]
\centering
\includegraphics[scale = 0.42]{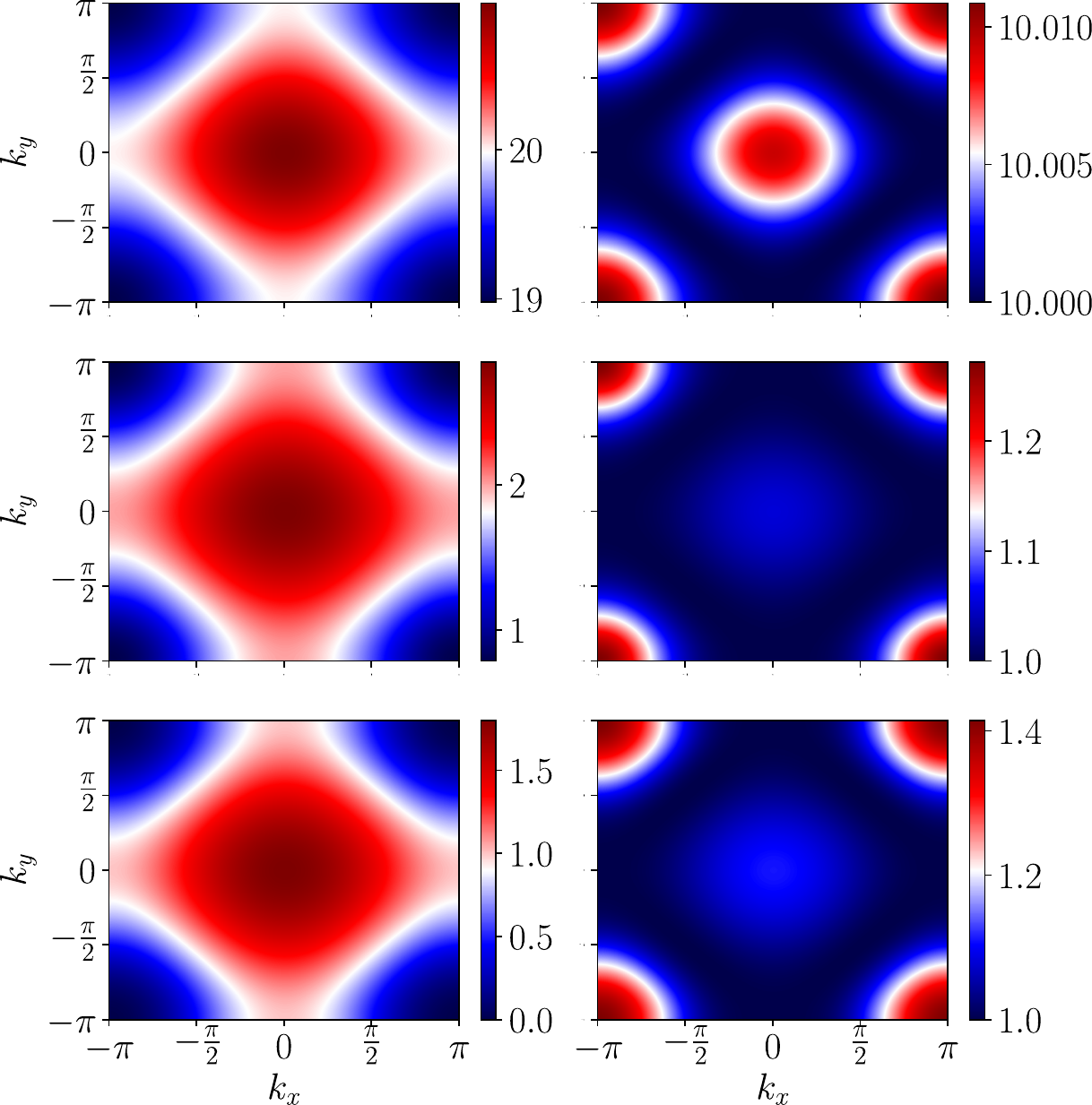}
\caption{\label{E_k_2D}
Analytical results concerning the complex excitation spectrum of the non-Hermitian transverse Ising model on a square lattice obtained via the bosonic approach relying on the
Holstein-Primakoff transformation together with the complex bosonic Bogolyubov transformation. The real and imaginary part of the excitation spectrum $E_{\bold{k}}$, 
denoted by $\Re(E_{\bold{k}})$ (first column) and $\Im(E_{\bold{k}})$ (second column) respectively, are plotted with respect to the
momentum $k_x$ and $k_y$, see Eq.~\eqref{spectrum}. The considered numerical parameters are (first line) $J=1$, $h=20$, $\gamma = 10$, (second line) $J=1$, $h=2$, $\gamma=1$, (third line) $J=h=\gamma=1$.}
\end{figure}

We develop the spin-wave theory of the model for the paramagnetic phase $J \ll h, \gamma$ via the Holstein-Primakoff transformation; we obtain a Hamiltonian that is quadratic and is best written in momentum space:
\begin{equation}
\hat{H} = \frac{1}{2} \sum_{\mathbf{k}} A_{\mathbf{k}} \left(\hat{a}^{\dag}_{\mathbf{k}} \hat{a}_{\mathbf{k}} + \hat{a}_{-\mathbf{k}} \hat{a}^{\dag}_{-\mathbf{k}} \right) + 
B_{\mathbf{k}} \left(\hat{a}^{\dag}_{\mathbf{k}} \hat{a}^{\dag}_{-\mathbf{k}} + \hat{a}_{\mathbf{k}} \hat{a}_{-\mathbf{k}} \right). 
\label{ham_2D_generic_form}
\end{equation}
Both momentum-dependent functions $A_{\bold{k}}$ and $B_{\bold{k}}$ are simple generalisation of the one-dimensional results:
\begin{subequations}
\begin{align}
& A_{\bold{k}} = \frac{J}{2}\left(\cos(\bold{k}\cdot\bold{x}) + \cos(\bold{k}\cdot \bold{y}) \right) + h + i \gamma, \label{A_k_2d} \\
& B_{\bold{k}} = \frac{J}{2}\left(\cos(\bold{k}\cdot \bold{x}) + \cos(\bold{k}\cdot \bold{y}) \right). \label{B_k_2d}
\end{align}
\end{subequations}
Using the complex bosonic Bogolyubov transformation defined in Eqs.~\eqref{b1} and \eqref{b2} and adapted to the two-dimensional case, the excitation spectrum
$E_{\bold{k}}$ of the non-Hermitian TFIM on a square lattice is given by 
\begin{equation}
E_{\bold{k}} = \mathrm{sgn}(\Re(A_{\bold{k}})) \sqrt{A_{\bold{k}}^2-B_{\bold{k}}^2},
\end{equation}
and is plotted on Fig.~\ref{E_k_2D} for different sets of parameters.
%

As in the 1D case, we recover the symmetry $\bold{k} \to -\bold{k}$ as well as the presence of a gap given by $h$ for the real part and $\gamma$ for the imaginary part when the 2D non-Hermitian quantum system is deep in the $z$ polarized phase i.e.~ $h \gg J$, see first line on 
Fig.~\ref{E_k_2D}. For the regime where $h \gtrsim J$, the gap in the imaginary part of $E_{\bold{k}}$ is still given by the dissipation strength $\gamma$ contrary to the one in the real part which is not given anymore by the transverse magnetic field $h$. Besides, on the third line of Fig.~\ref{E_k_2D}, due to the specific value of the parameters, the momentum-dependent function $\mathrm{sgn}(\Re(A_{\bold{k}}))$ in the expression of $E_{\bold{k}}$ is not defined at the four points corresponding to the edges of the first Brillouin zone and given by $\bold{k} = (-\pi,-\pi), (-\pi,\pi), (\pi,-\pi), (\pi,\pi)$. Furthermore, for the same set of parameters, we notice that there is a gap closing for the real part of the excitation spectrum $\Re(E_{\bold{k}})$ meaning that the latter is gapless. This leads to a breakdown of the bosonic approach based on the linear spin-wave theory (truncation of Holstein-Primakoff transformation together with the complex bosonic Bogolyubov transformation). 

The generalization to a hypercubic lattice of dimension $D$ is straightforward. Indeed, the excitation spectrum is still given by the following equation 
\begin{equation}
E_{\bold{k}} = \mathrm{sgn}(\Re(A_{\bold{k}})) \sqrt{A_{\bold{k}}^2-B_{\bold{k}}^2},
\end{equation}
where $\bold{k} = (k_1~ k_2 ~... ~k_{D-1}~ k_D)^T$ is a $D$-dimensional vector and where the momentum-dependent functions $A_{\bold{k}}$ and $B_{\bold{k}}$ are given by
\begin{subequations}
\begin{align}
& A_{\bold{k}} = h + i\gamma + \frac{J}{2} \sum_{i=1}^{D} \cos(\bold{k}\cdot \bold{d}_i),  \\
& B_{\bold{k}} = \frac{J}{2} \sum_{i=1}^{D} \cos(\bold{k}\cdot \bold{d}_i),
\end{align}
\end{subequations}
\noindent
with $\bold{d}_1 = (a_1~0~...~0)^T$, $\bold{d}_2 = (0~ a_2~ 0~...~0)^T$, $\bold{d}_j = (0~ ...~0~ a_j~0~ ... ~0)^T$, $\bold{d}_D = (0~...~0~a_D)^T$
and where $a_i = a = 1$ with $i \in [|1,D|]$.

\section{Spin-wave theory for the Quench Dynamics}
\label{sec:dynamics}

We now investigate the out-of-equilibrium dynamics of the non-Hermitian TFIM discussing a sudden global quench starting from an initial state $\ket{\Psi_0}$
corresponding to the ground state of the Hermitian TFIM ($\gamma = 0$) deep in the paramagnetic phase ($h \gg J$). 
At time $t=0$ we switch on the dissipative coupling ($\gamma\neq 0$) and we study the time evolution with the post-quench non-Hermitian Hamiltonian $\hat{H}$ of the TFIM.
Both pre- and post-quench Hamiltonians are confined relatively deep into the paramagnetic phase where the spin-wave theory can be safely employed, at least for the spectrum. As we will see this is not necessary the case for the long-time dynamics. The initial state can be roughly thought of as a fully-polarised product state $ \ket{\uparrow \uparrow ... \uparrow \uparrow}$, although in the calculations we will use the correct expression, i.e.~ the ground-state of the Hermitian TFIM ($\gamma = 0$) for $h=5J$ where $J$ will be fixed to unity.
The time-evolved state reads:
\begin{equation}
\ket{\Psi(t)} = \frac{e^{-i\hat{H}t}\ket{\Psi_0}}{||e^{-i\hat{H}t}\ket{\Psi_0}||}, 
\label{psi_t}
\end{equation}
and the normalization factor is essential since $\hat{H}$ is non-Hermitian and reduces the norm of the state, thus making the dynamics a non-linear function of the state.

Let us begin our discussion considering the case $J=0$ where all lattice sites are decoupled and focus on the single site dynamics which can be solved exactly. 
For a generic initial state $\ket{\Psi_0} = \alpha \ket{\uparrow} + \beta \ket{\downarrow}$, with $|\alpha|^2+ |\beta|^2=1$, the time-evolved state reads:
\begin{equation}
 \ket{\Psi(t)} = \displaystyle \frac{\alpha e^{\frac{iht}{2}}e^{-\frac{\gamma t}{2}} \ket{\uparrow} + \beta e^{-\frac{iht}{2}}e^{\frac{\gamma t}{2}}\ket{\downarrow}}{\sqrt{|\alpha|^2 e^{-\gamma t} + |\beta|^2 e^{\gamma t}}},
\end{equation}

\noindent
from which it follows that, as along as $\alpha$ and $\beta$ are different from zero, the asymptotic state is $\ket{\downarrow}$ for $\gamma>0$ and $\ket{\uparrow}$ for $\gamma<0$.
This is for instance reflected by the time-evolution of the projector on the $\ket{\downarrow}$ state:

\begin{equation}
  \langle \Psi(t)| \hat{S}^- \hat{S}^+ | \Psi(t) \rangle = \frac{|\beta|^2}{|\beta|^2+|\alpha|^2 e^{-2\gamma t}  }.
 \label{corr}
\end{equation}

The goal of the next sections is to use the spin-wave theory developed so far in order to discuss the dynamics of the spin excitations in the system when $J$ is different from zero.

Before dealing with the calculations, we can already anticipate that our approach will not be able to describe the non-unitary dynamics at all times for $\gamma > 0$.
Since the linear spin-wave theory is a good approximation when the number of bosons $\langle \hat a_k^\dagger \hat a_k \rangle \ll 1$, the spin state must be close to the paramagnetic state $\ket{\uparrow \uparrow ... \uparrow \uparrow}$.
Yet, as we have just seen, for $\gamma > 0$ the long-time dynamics drives the system towards the state $\ket{\downarrow...\downarrow}$ where the spin-wave theory is not supposed to work. The study that we are going to perform is thus expected to give a good quantitative description only at short times. We will discuss this point extensively. 
On the other hand, we expect this problem not to arise for $\gamma<0$ because in the long-time limit the system remains close to the state $\ket{\uparrow \uparrow ... \uparrow \uparrow}$ which is well described in terms of a small number of bosonic excitations, thus maintaining the system in a regime where spin-wave theory works.

\subsection{One-dimensional lattice: equations of motion and quench dynamics}
\label{1d_case}

In order to characterize the quench dynamics of our model, we derive the equations of motion (EoM) associated to the quadratic bosonic correlators:
\begin{equation}
 G_k(t) = \langle \hat{a}^{\dag}_k \hat{a}_k \rangle_t,
 \qquad 
 F_k(t) = \langle \hat{a}_k \hat{a}_{-k} \rangle_t,
\end{equation}
where
$\langle ... \rangle_t = \langle \Psi(t) | ... | \Psi(t) \rangle$ with $\ket{\Psi(t)}$ given at Eq.~\eqref{psi_t}. 
Note that $F_k(t)^* =\langle \hat{a}^{\dag}_{-k} \hat{a}^{\dag}_k\rangle_t = \langle \hat{a}^{\dag}_k \hat{a}^{\dag}_{-k} \rangle_t$ and
$\langle \hat{a}^{\dag}_k \hat{a}_q \rangle_t = \delta_{k,q}G_k(t)$ due to the momentum conservation. 
We obtain the two non-linear coupled differential equations (see Appendix~\ref{EMT} for details on the derivation): 
\begin{widetext}
\begin{subequations}
\label{diff_eq_G_k_t}
\begin{align}
& \frac{\mathrm{d}}{\mathrm{d}t}F_k(t) = 4\Im(A_k)F_k(t)G_k(t) - 2iA_kF_k(t) -iB_k \left(1+2G_k(t)\right); \label{diff_eq_F_k_t} \\
& \frac{\mathrm{d}}{\mathrm{d}t}G_k(t) = -2B_k\Im(F_k(t)) + 2\Im(A_k)\left(|F_k(t)|^2 + G_k(t) + G_k(t)^2 \right) \label{diff_eq_G}. 
\end{align}
\end{subequations}
\end{widetext}

\noindent
These equations of motion are valid for any value of $\gamma$, be it positive or negative. In the rest of this section we will focus on the case $\gamma>0$, where the breakdown of spin-wave theory is expected, and postpone the discussion of the case $\gamma<0$ to Sec.~\ref{Sec:QuenchNegativeGamma}.


\begin{figure}[h!]
\includegraphics[scale = 0.46]{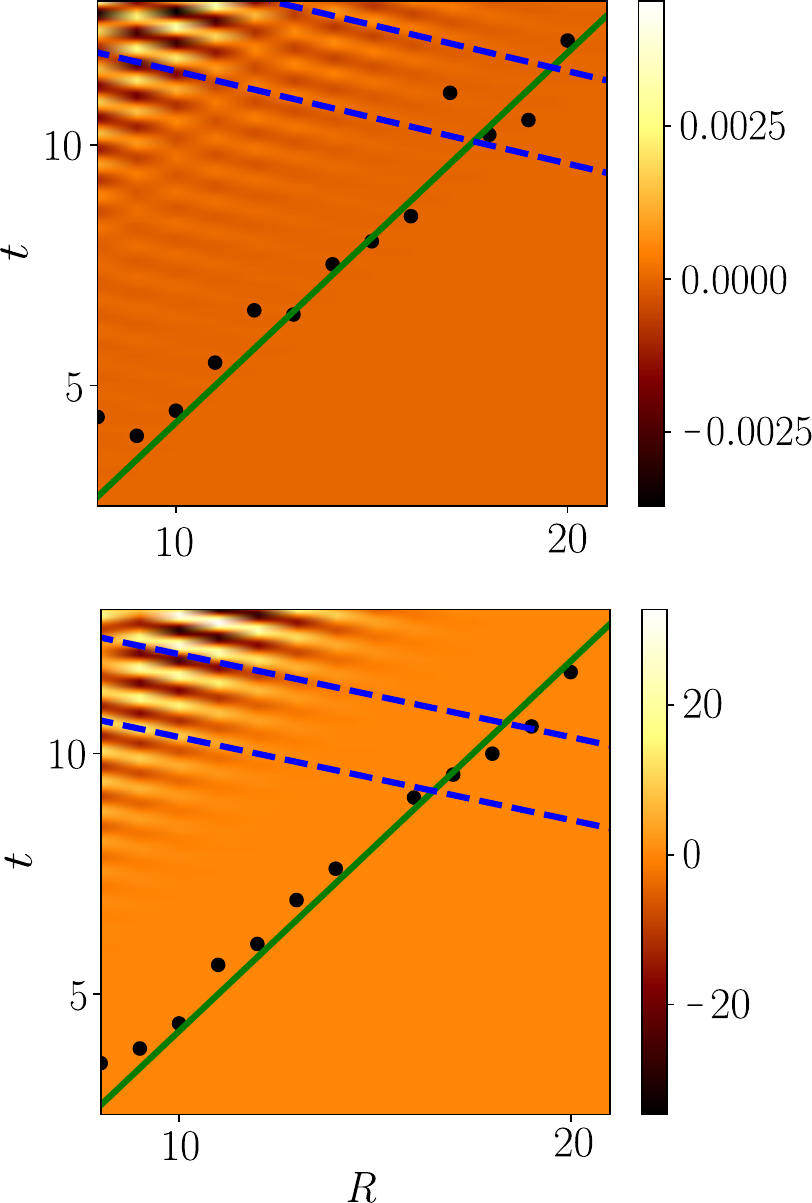}
\caption{Space-time plot of the one-body correlation function $G_R(t)$ for a sudden global quench on $\gamma$ of the TFIM in the paramagnetic phase. 
We start from the ground state at $\gamma = 0$ (Hermitian case) and evolve it in time with $\gamma > 0$ (non-Hermitian case) for the transverse Ising chain in the $z$ polarized phase.
(Top) The results are obtained by integrating the EoM in Eq.~\eqref{diff_eq_G_k_t}. (Bottom) Theoretical guess of the correlation function characterised by Eq.~\eqref{G_R_t_th_guess}.
The parameters of the simulation are: $N_s = 200$, $h=5$, $\gamma = 0.2$, $J=1$.
The slope of the superimposed lines are: $V_{\mathrm{m}} = -5.2$ and $V_{\mathrm{CE}} = 1.3$ (top panel) and $V_{\mathrm{m}} = -5.8$ and $V_{\mathrm{CE}} = 1.3$ (bottom panel).}
\label{plot2}
\end{figure}


We now focus on the spatial propagation of the quantum correlations induced by the dissipative quench by analyzing the bosonic one-body density matrix:
\begin{align}
G_R(t) =& \langle \hat{a}^{\dag}_R \hat{a}_0 \rangle_t =  
\frac{1}{N_s} \sum_k \cos(kR)G_k(t), 
\label{G_R_t_G_k_t}
\end{align}
that should approximate the $\langle \hat{S}^{-}_R \hat{S}^{+}_0 \rangle_t $ spin correlator.
Fig.~\ref{plot2} presents the results of our numerical simulations, where we clearly notice the emergence of a light-cone structure in the space-time plot of $G_R(t)$~\cite{lieb1972}. 
A clear correlation edge (CE) separates the causal region where the correlations have non-zero values from the non-causal region where the quantum correlations are zero.
They propagate ballistically with a velocity $V_{\mathrm{CE}}$ which can be deduced from the slope of the CE, highlighted by a green solid line in Fig.~\ref{plot2}. In Appendix~\ref{ce_track_appendix} we describe several techniques to numerically estimate $V_{\rm CE}$. 

Moreover, we observe that in the vicinity of the CE the $G_R(t)$ displays a series of local minima and maxima which also propagate ballistically, highlighted in Fig.~\ref{plot2}(top) by several blue dashed lines. 
This behavior of the space-time correlations for a non-Hermitian short-range interacting quantum lattice model is reminiscent of that of 
Hermitian short-range and long-range interacting quantum lattice models in the so-called local regime~\cite{despres2018, despres2019, despres2021}.
This regime refers to large values of $\alpha$ which corresponds to the power-law exponent of the long-range algebraic interactions. 

In order to give a clearer interpretation of the space-time pattern of $G_R(t)$, we present a simplified theoretical model based on the following analytical expression, see Appendix \ref{th_guess} for more details,

\begin{align}
& G_R(t) = \frac{1}{N_s} \sum_k F_k \cos(kR) \cos(2\Re(A_k)t)\exp(2\Im(A_k)t), 
\label{G_R_t_th_guess}
\end{align}

\noindent
where $A_k = h + (J/2)\cos(k) + i\gamma$ and the momentum-dependent function $F_k$ corresponding to the sum of three Gaussians is given by

\begin{align}
 F_k =& \exp \left(-\frac{(k-\pi)^2}{2} \right) + \exp \left(-\frac{(k+\pi)^2}{2} \right) \nonumber \\
& + \exp \left(-\frac{k^2}{2} \right),
\end{align}

\noindent
where $k \in [-\pi,\pi]$. Note that the numerical space-time pattern of $G_R(t)$ deduced from the EoM, see top panel on Fig.~\ref{plot2}, is qualitatively very well reproduced by our theoretical guess, see bottom panel on Fig.~\ref{plot2}. Indeed, the twofold linear structure as well as the two characteristic velocities $V_{\mathrm{CE}}$ and $V_{\mathrm{m}}$ are well captured.

Concerning the long-time quench dynamics of our model, we find a breakdown of the linear spin-wave theory. By looking at the expressions of the EoM we could anticipate an exponential growth of  $G_k(t)$ as a function of the time $t$, since its time derivative is proportional  to $\Im(A_k) = \gamma$,
see Eq.~\eqref{diff_eq_G_k_t}. This brings quickly the system to a regime where the linear approximation breaks-down. For instance, for the set of parameters considered in Fig.~\ref{plot2},
we find a value of $\mathrm{max}(G_k(t)) = \mathrm{max} (\langle a^{\dag}_k a_k \rangle_t)$ significantly lower than $2s = 1$. Hence, for $t \leq t_{\mathrm{f}}$ where $t_{\mathrm{f}}$ refers
to the observation time, the linear spin-wave approximation is well justified. However, it is not the case anymore when $t > t_{\mathrm{f}}$ because of $\mathrm{max}(G_k(t))\gtrsim 2s$. 
In addition to this exponential growth we find a genuine divergence of $G_k(t)$, on a time scale which is controlled by $1/\gamma$. In order to rationalize this surprising result in the next section we will simplify the problem and study the dynamics of a single bosonic mode, corresponding to selecting only $k=0$ momentum in the spin wave theory.

\subsection{Dynamical Instability in a single-mode problem}

To understand the origin of the dynamical instability discussed at the end of previous section we now investigate the single-mode $k=0$ problem, which is the simplest situation where we can
test the predictions of the spin-wave EoM developing a comparison with the numerics based on exact diagonalization (ED). 
This study will provide a simple intuition concerning the increase of $G(t) = \langle \hat{a}^{\dag} \hat{a}
\rangle_t$ and thus the breakdown of the linear spin-wave approximation at large times, which makes the spin-wave theory useless.
In fact, we will observe that the equations predict a mathematical singularity at finite times, a problem that will plague all our studies of dynamics using spin-wave theory when $\gamma > 0$.

\begin{figure}[t]
\includegraphics[scale = 0.34]{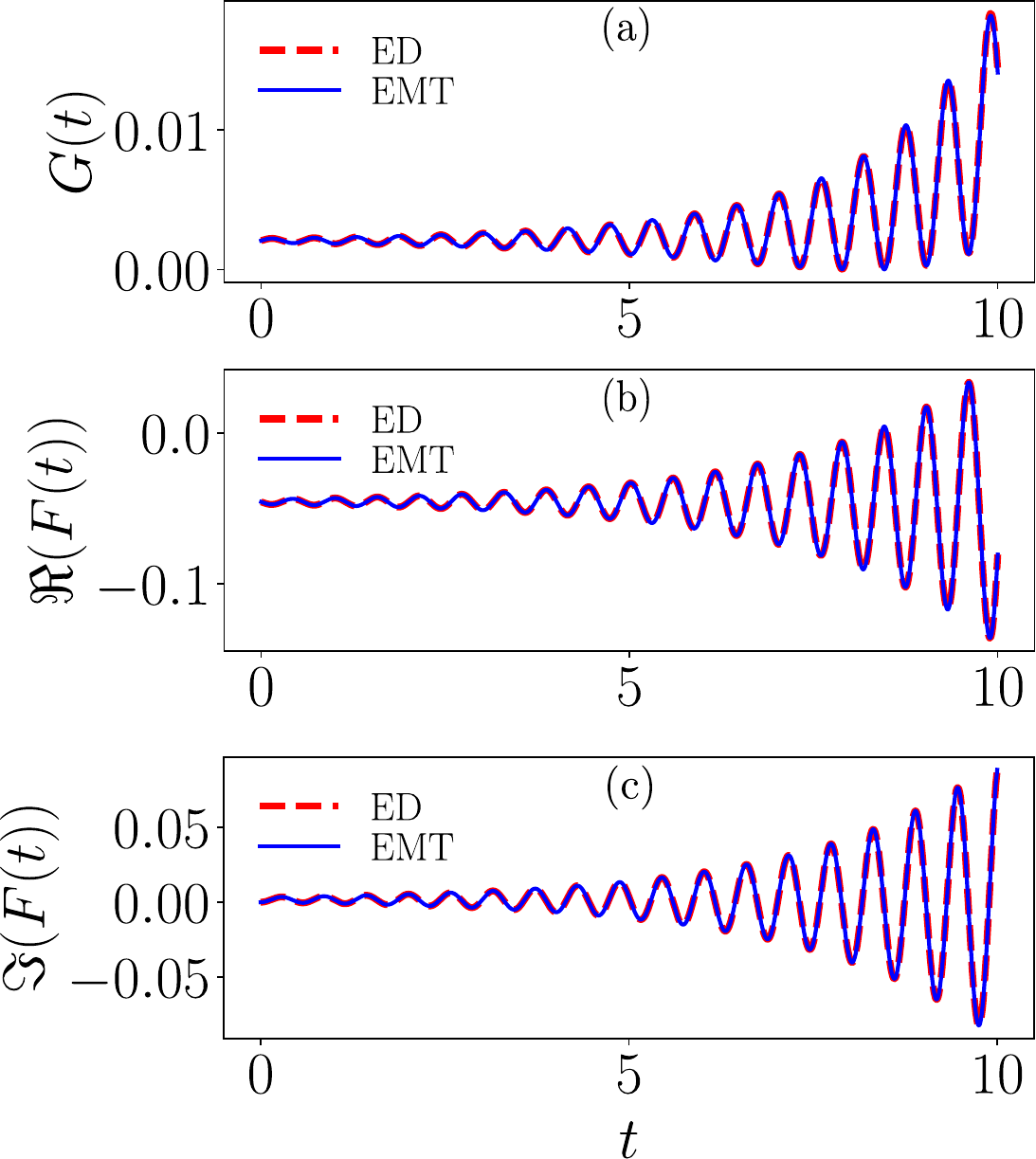}
\caption{Single-mode two-body correlators as a function of time $t$ where a sudden global quench has been considered on $\gamma$ from $\gamma = 0$ to $\gamma \neq 0$ for the non-Hermitian  transverse Ising chain confined in the $z$ polarized phase. (a) $G(t) = \langle \hat{a}^{\dag} \hat{a} \rangle_t$, (b) $\Re(F(t)) = \Re(\langle \hat{a}\hat{a} \rangle_t)$, (c) $\Im(F(t)) = \Im(\langle \hat{a}\hat{a} \rangle_t)$. For each plot, we compare the results found using exact diagonalization (see dashed red lines) and using the equation of motion technique (see solid blue lines). Numerical parameters : $h=5$, $\gamma = 0.2$, $J=1$, $N_{\mathrm{max}}=10$.}
\label{plot_single_mode}
\end{figure}

Let us begin with the benchmark of the EoM of spin-wave theory agains ED; the single-mode Hamiltonian of the non-Hermitian transverse Ising chain is given by
\begin{equation}
 \hat{H} = \frac{A}{2} \left(\hat{a}^{\dag} \hat{a} + \hat{a}\hat{a}^{\dag} \right) 
+ \frac{B}{2} \left( \hat{a}^{\dag} \hat{a}^{\dag} + \hat{a} \hat{a} \right), 
\label{H_single_mode}
\end{equation}
with $A = h + (J/2) + i\gamma$, $B = J/2$. $\hat{a}$ is a short-hand notation for $\hat{a}_{k=0}$.
We specify the EOMs in~\eqref{diff_eq_G_k_t} to this case and integrate them numerically using a fourth-order Runge-Kutta method. 
Similarly, we compute an ED study of the dynamics of Hamiltonian~\eqref{H_single_mode} truncating the Hilbert space at the maximal number of bosons $N_{\rm max}$.
In both cases, the initial state corresponds to the ground state of the Hamiltonian~\eqref{H_single_mode} for $\gamma = 0$.
The results obtained with both methods are compared in Fig.~\ref{plot_single_mode}, where we observe a perfect agreement as long as we consider low bosonic occupation numbers, $G(t) \ll 1$, here of the order of $10^{-2}$. 

We now inspect more closely the properties of the solution obtained integrating the bosonic EoM.
We study an initial state $\ket{\Psi_0} = \ket{\Psi(0)}$ that is a single-mode squeezed vacuum (SMSV) state i.e.~$\ket{\Psi_0} = \ket{\mathrm{SMSV}} = \hat S(\xi) \ket{0}$, where $\hat S(\xi)$
corresponds to the squeezing operator and $\xi = r e^{i\phi}$ is the squeezing parameter. 
It corresponds to the generic ground state of the Hamiltonian in Eq.~\eqref{H_single_mode}
for $\gamma = 0$ and can be expanded as 
\begin{equation}
\ket{\Psi_0} = \frac{1}{\sqrt{\cosh(r)}} \sum_{n=0}^{+\infty} \left(-e^{i\phi}\tanh(r) \right)^n \frac{\sqrt{(2n)!}}{2^n n!} \ket{2n},
\end{equation}
on the bosonic Fock basis.
Since we are interested in the effect of the local dephasing noise on the quench dynamics, we focus on the simple model
\begin{equation}
\hat{H}_s = i \frac{\gamma}{2} \hat{a}^{\dag}\hat{a}, 
\end{equation}
where the prefactor $1/2$ is for convenience. 
It corresponds to the anti-Hermitian part of the Hamiltonian~\eqref{H_single_mode}.

With an explicit calculation, one easily obtains that the non-normalised state $\ket{\tilde \Psi (t)} = e^{-i \hat{H}_s t} \ket{\Psi_0}$ reads
\begin{equation}
\ket{\tilde \Psi (t)} = \frac{1}{\sqrt{\cosh(r)}} \sum_{n=0}^{+\infty} \left(-e^{i\phi}\tanh(r)e^{\gamma t} \right)^n \frac{\sqrt{(2n)!}}{2^n n!} \ket{2n}.
\end{equation}
After some analytical calculations, we obtain an expression for its norm:
\begin{equation}
||\ket{\tilde{\Psi}(t)}||^2 = \frac{1}{\cosh(r)} \sum_{n=0}^{+\infty} \left( \frac{\tanh(r)e^{\gamma t}}{2} \right)^{2n} \frac{(2n)!}{(n!)^2},
\end{equation}
which can be made more readable using the Stirling approximation $n! \sim n^n$ for large $n$:
\begin{equation}
||\ket{\tilde{\Psi}(t)}||^2 = \frac{1}{\cosh(r)} \sum_{n=0}^{+\infty} \left(\tanh(r)e^{\gamma t}\right)^{2n}.
\end{equation}
The latter converges only if $\tanh(r)e^{\gamma t} < 1$ where $ 0 < \tanh(r) < 1$ since $r > 0$. The previous inequality implies the existence of a typical time 
\begin{equation}
t_{\mathrm{f}} \approx \frac{1}{\gamma} \ln \left(\frac{1}{\tanh(r)} \right),
\label{tf}
\end{equation}
at which the bosonic theory encounters a singularity that
is responsible for the divergence of 
$G(t)$.
Not only thus the quadratic bosonic theory is problematic because it cannot describe the physical fact that $0 \leq \langle S^- S^+ \rangle_t \leq 1$, but it also predicts that $G(t)$ diverges at finite time. We anticipate here that the divergence at finite time has been observed in all the numerical simulations of the time evolution of the linear spin-wave theory performed in this study
with $\gamma > 0$. In order to have a large value of $t_{\rm f}$ we have always used a relatively small value of $\gamma$, as detailed in the captions of the figures.
We emphasize that this divergence is not cured by including non-linear terms in the bosonic field, see Appendix~\ref{nl}.



\subsection{Square lattice}

\begin{figure*}[ht]
\includegraphics[scale = 0.28]{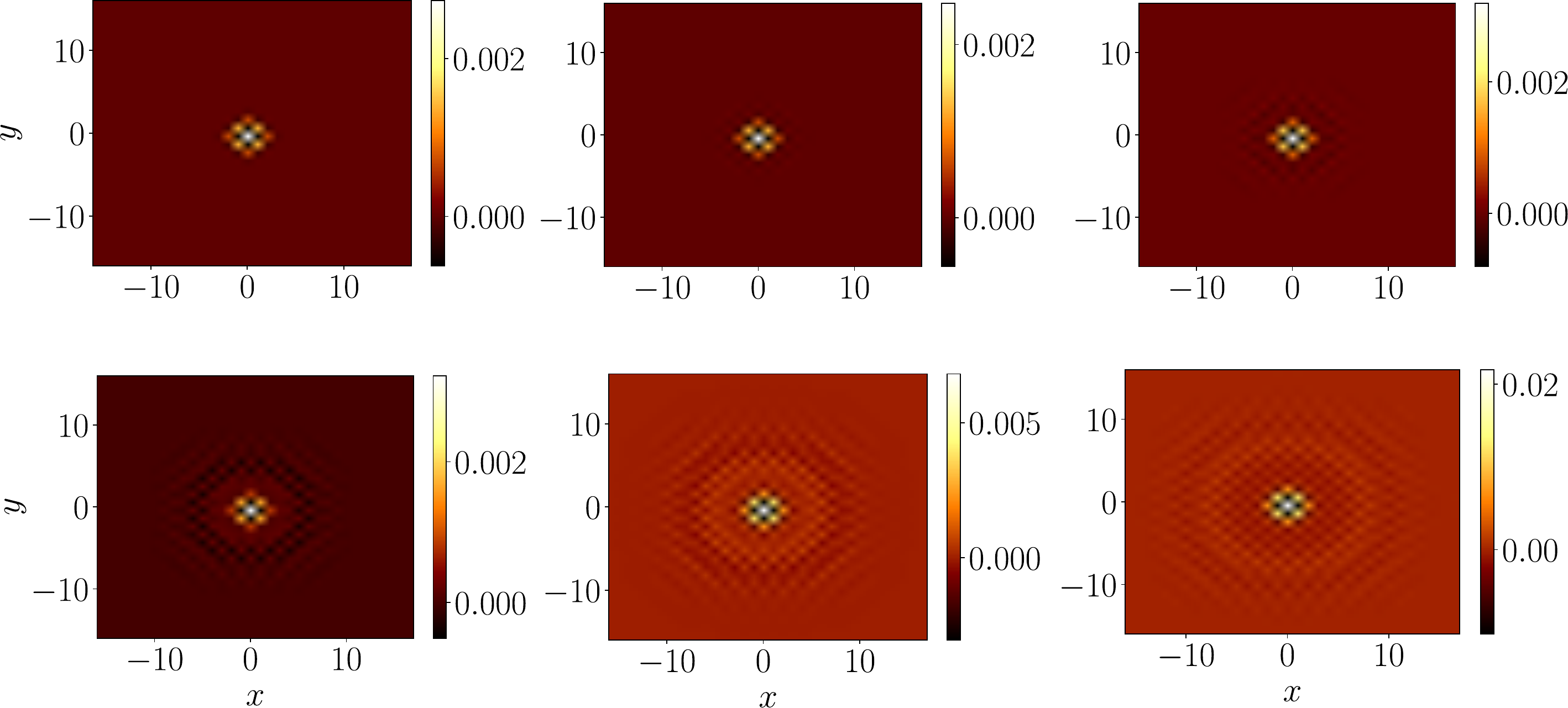}
\caption{One-body correlation function $G_{\mathbf{R}}(t) = G_{x,y}(t)$ for a sudden global quench on $\gamma$ from $\gamma = 0$ (Hermitian case) 
to $\gamma \neq 0$ (non-Hermitian case) for the transverse Ising model on a square lattice and confined in the $z$ polarized phase.
Numerical results found using the equation of motion technique to deduce $G_{\mathbf{k}}(t)$, see Eq.~\eqref{G_R_t_G_k_t_square_lattice}.
The parameters are : $N_s = 100$, $h=5$, $\gamma = 0.2$, $J=1$, $t_{\mathrm{i}}=0$, $t_{\mathrm{f}} = 10$, $\mathrm{steps} = 300$. (first line from left to right)
$t = t_{\mathrm{i}} = 0$, $t = t_{\mathrm{f}}/5 = 2$, $t = 2t_{\mathrm{f}}/5 = 4$ (second line from left to right) $t = 3t_{\mathrm{f}}/5 = 6$, $t = 4t_{\mathrm{f}}/5 = 8$, 
$t = t_{\mathrm{f}} = 10$.}
\label{one_body_corr_fct_square_lattice}
\end{figure*}

We now turn on the two-dimensional case, i.e.~the square lattice. 
Using the notation employed in Sec.~\ref{Sec:Exc:Sp:2D}, we introduce the functions $G_{\bold{k}}(t) = \langle \hat{a}^{\dag}_{\mathbf{k}} \hat{a}_{\mathbf{k}} \rangle_t$ and $F_{\bold{k}}(t) = \langle \hat{a}_{\bold{k}} \hat{a}_{-\bold{k}} \rangle_t$ 
which obey the same EoM derived for the one-dimensional case in Eq.~\eqref{diff_eq_G_k_t}. The one-body density matrix reads
\begin{equation}
G_{\mathbf{R}}(t) = \frac{1}{N_s^2} \sum_{\mathbf{k}} \cos(\mathbf{k}\cdot\mathbf{R}) G_{\mathbf{k}}(t),
\label{G_R_t_G_k_t_square_lattice}
\end{equation}
where $\mathbf{R} = (x ~~ y)^T$ and is plotted at different times in Fig.~\ref{one_body_corr_fct_square_lattice} for a sudden global quench from $\gamma = 0$ to $\gamma \neq 0$.

From Fig.~\ref{one_body_corr_fct_square_lattice} dealing with the short-time quench dynamics of our 2D model, we clearly notice that correlations propagate in the square lattice from the central lattice site with time. This pattern is characterized an approximately circular CE, whose radius increases linearly with time. The velocity can be extracted by fitting the behaviour of the system both along the $x$ and $y$ direction.
We track $x^*$ and $y^*$ corresponding to the maximal value along the $x$ and $y$ direction of a 
non-zero correlation for several values of the time $t$. The data are extracted from Fig.~\ref{one_body_corr_fct_square_lattice} and we find $V_{\rm CE} \simeq 1.1$, see Appendix~\ref{cev}
for another numerical technique to determine the correlation edge velocity $V_{\mathrm{CE}}$. \\

\subsection{Fermionic approach for the one-dimensional lattice}

In what follows, we consider a fermionic approach to deal with the quench dynamics of the non-Hermitian transverse Ising chain confined in the $z$ polarized phase. This will permit to compare its predictions to those obtained by using the bosonic approach discussed at Subsection \ref{1d_case}. Similarly as the bosonic approach, the fermionic one consists in calculating analytically the equations of motion associated to the two-body correlators, also called second-order or quadratic correlators, expressed in terms of the fermionic annihilation and creation operators denoted by $\hat{c}_k$ and $\hat{c}^{\dag}_k$. The two interesting correlators are $\mathcal{G}_k(t) = \langle \hat{c}^{\dag}_k \hat{c}_k \rangle_t$ and $\mathcal{F}_k(t) = \langle \hat{c}_k \hat{c}_{-k} \rangle_t$. $\langle ... \rangle_t = \langle \Psi(t) | ... | \Psi(t) \rangle$ and $\ket{\Psi(t)}$ is given at Eq.~\eqref{psi_t} where the Hamiltonian 
$\hat{H}$ is defined at Eq.~\eqref{H_fermionic_quadratic}. We obtain the two following complex coupled differential equations~\cite{turkeshi2023,legal2023}

\begin{widetext}
\begin{align}
& \frac{\mathrm{d}}{\mathrm{d}t}\mathcal{F}_k(t) = -8 \Im(\mathcal{A}_k)\mathcal{F}_k(t)\mathcal{G}_k(t) - 4i\mathcal{A}_k\mathcal{F}_k(t) + 2i\mathcal{B}_k(1-2\mathcal{G}_k(t)), \label{fermion_diff_eq_fkt} \\
& \frac{\mathrm{d}}{\mathrm{d}t}\mathcal{G}_k(t) = 4\Im(\mathcal{A}_k)(|\mathcal{F}_k(t)|^2 + \mathcal{G}_k(t) - \mathcal{G}_k(t)^2 ) + 4i\mathcal{B}_k \Re(\mathcal{F}_k(t)), \label{fermion_diff_eq_gkt}
\end{align}
\end{widetext}

\noindent
which can be solved numerically using a fourth-order Runge-Kutta method, see Appendix \ref{EMT_fermions} for more details about the calculation of the equations of motion. In the following, two different observables, namely the spin-spin correlation function denoted by $C^{zz}_R(t)$ and defined as 

\begin{equation}
C^{zz}_R(t) = \langle \hat{S}^z_R \hat{S}^z_0 \rangle_t - \langle \hat{S}^z_R \rangle_t \langle \hat{S}^z_0 \rangle_t,
\label{czzrt}
\end{equation}

\noindent
and the local magnetization $\langle \hat{S}^z_R \rangle_t$, are discussed. \\

\paragraph{Local magnetization -} We now turn to the second observable which is the local magnetization $\langle \hat{S}^z_R \rangle_t$. Using a similar procedure as the one used previously for the spin-spin correlations along the $z$ axis, we find for the bosonic approach the following analytical expression

\begin{equation}
\langle \hat{S}^z_R \rangle_t = \frac{1}{2} - \frac{1}{N_s} \sum_k G_k(t),
\label{szrt_boson}
\end{equation}

\noindent
and for the fermionic approach, we obtain a very similar form given by 

\begin{equation}
\langle \hat{S}^z_R \rangle_t = \frac{1}{2} - \frac{1}{N_s} \sum_k \mathcal{G}_k(t).
\label{szrt_fermion}
\end{equation}

\begin{figure}[t]
\includegraphics[scale = 0.45]{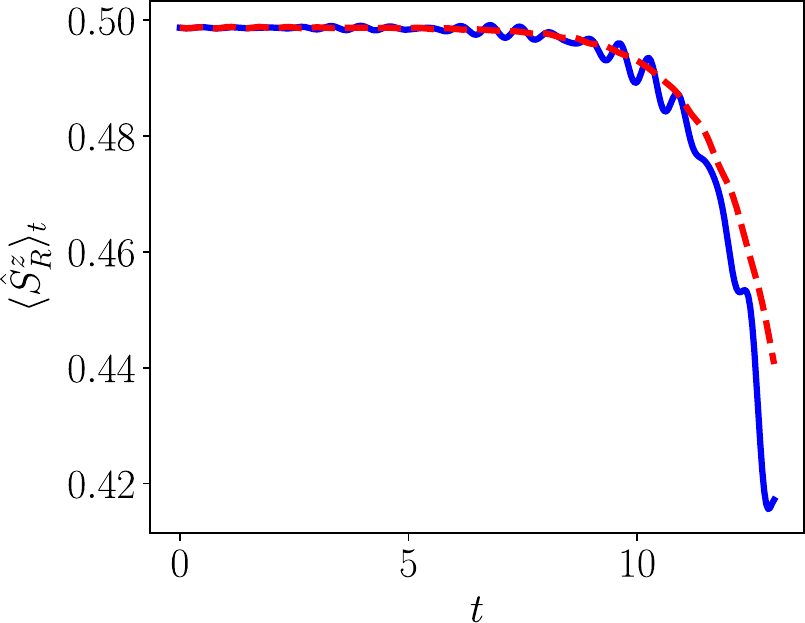}
\caption{Local magnetization $\langle \hat{S}^z_R\rangle_t$ for a sudden global quench on $\gamma$ from $\gamma = 0$ to $\gamma \neq 0$ for the transverse Ising chain in the $z$ polarized phase. For the profile associated to the bosonic approach, see solid blue line 
and Eq.~\eqref{szrt_boson} and for the one associated to the fermionic approach, see dashed red line and Eq.~\eqref{szrt_fermion}. The parameters are : $N_s = 200$, $h=5$, $\gamma = 0.2$, $J=1$, $t_{\mathrm{i}}=0$, $t_{\mathrm{f}} = 13$, $\mathrm{steps} = 300$.}
\label{plot_lm}
\end{figure}

\noindent
Note that both theoretical expressions of the local magnetization are independent of the distance $R$ due to the translational invariance of the 
Hamiltonian associated to the transverse-field Ising chain. The theoretical expression associated to the time-dependent local magnetization using the
bosonic and the fermionic approach is plotted in Fig.~\ref{plot_lm} for the transverse-field Ising chain confined in the $z$ polarized phase where a 
sudden global quench on $\gamma$ the dissipation strength is considered. We can clearly see a quantitative agreement between both expressions at short, intermediate and large times where the maximal relative error is very small. This means that the bosonic and fermionic approaches provide very similar 
results as expected and thus are equivalent even if the transverse-field Ising chain is confined not very deep into the $z$ polarized phase. Note that 
at time $t=0$, the local magnetization is close to $1/2$ due to the fact that the initial Hamiltonian is confined relatively deep into the paramagnetic phase.

\paragraph{Spin-spin correlation function -} Concerning the spin-spin correlation function $C^{zz}_R(t)$, the latter can be expressed in terms of the two-body correlators $F_k(t)$ and $G_k(t)$
($\mathcal{F}_k(t)$ and $\mathcal{G}_k(t)$) when the bosonic (fermionic) approach is considered. The procedure to follow in order to deduce the
analytical expression of $C^{zz}_R(t)$ is very similar for the two theoretical approaches. Indeed, the first step consists in using the Holstein-Primakoff
(Jordan-Wigner) transformation. The second step is to use the bosonic (fermionic) Wick theorem for the bosonic (fermionic) creation and annihilation operators. Then, we rely on the Fourier transform in order to manipulate the operators in momentum space and finally the momentum conservation is used. For the bosonic approach, the theoretical expression of $C^{zz}_R(t)$ is given by 

\begin{equation}
C^{zz}_R(t) = \frac{1}{N_s^2} \sum_{k,k'} e^{i(k-k')R} (F_k(t)^* F_{k'}(t) + G_k(t)(1 + G_{k'}(t))),
\label{czzrt_boson}
\end{equation}

\noindent
where $F_k(t) = \langle \hat{a}_k \hat{a}_{-k} \rangle_t$ and $G_k(t) = \langle \hat{a}^{\dag}_k \hat{a}_k \rangle_t$. Concerning the fermionic approach,
the spin-spin correlation function $C^{zz}_R(t)$ has the following expression 

\begin{equation}
C^{zz}_R(t) = \frac{1}{N_s^2} \sum_{k,k'} e^{i(k-k')R} (\mathcal{F}_k(t)^* \mathcal{F}_{k'}(t) + \mathcal{G}_{k}(t)(1-\mathcal{G}_{k'}(t))),
\label{czzrt_fermion}
\end{equation}

\noindent
with $\mathcal{F}_k(t) = \langle \hat{c}_k \hat{c}_{-k} \rangle_t$ and $\mathcal{G}_k(t) = \langle \hat{c}^{\dag}_k \hat{c}_k \rangle_t$.

\begin{figure}[t]
\includegraphics[scale = 0.47]{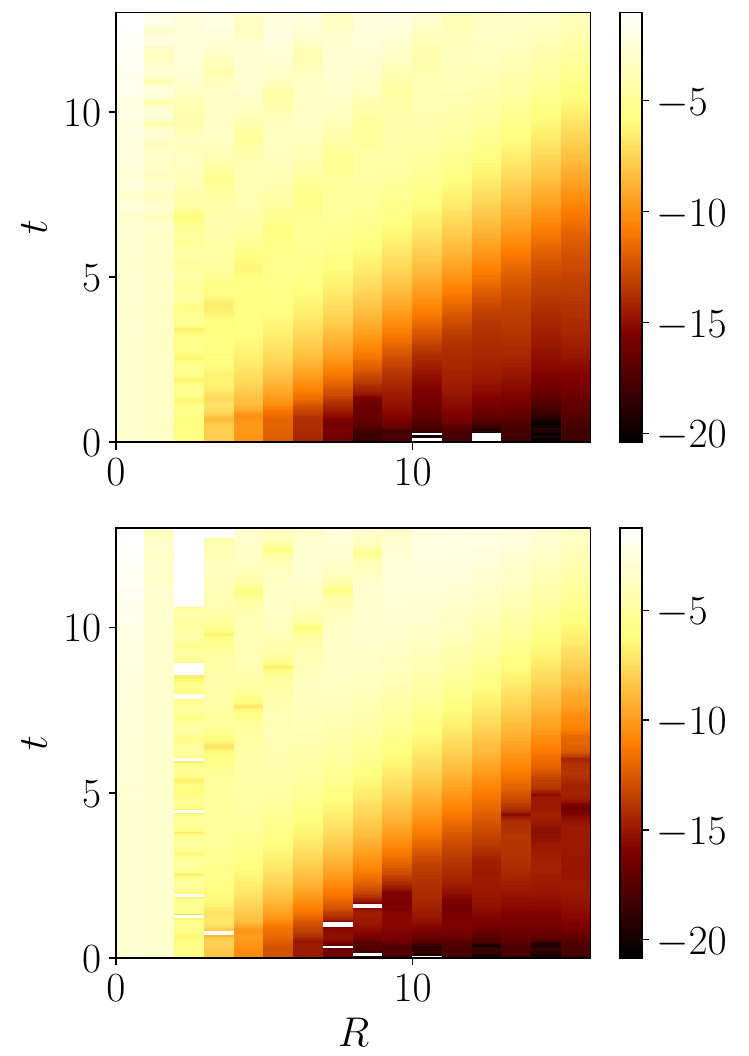}
\caption{Space-time spin-spin correlation function $C^{zz}_R(t)$ defined at Eq.~\eqref{czzrt} for a sudden global quench on $\gamma$ from $\gamma = 0$ to $\gamma \neq 0$ for the transverse Ising chain in the $z$ polarized phase. Real part of $C^{zz}_R(t)$
when using the bosonic approach, see Eq.~\eqref{czzrt_boson} (top panel), real part of $C^{zz}_R(t)$ when considering the fermionic approach, see
Eq.~\eqref{czzrt_fermion} (bottom panel). The parameters are : $N_s = 200$, $h=5$, $\gamma = 0.2$, $J=1$, $t_{\mathrm{i}}=0$, $t_{\mathrm{f}} = 13$, $\mathrm{steps} = 300$, $R_{\mathrm{i}} = 0$, $R_{\mathrm{f}} = 15$. Note that the logarithmic value of the real part of $C^{zz}_R(t)$ is considered and denoted
by $\mathrm{log}_{10}(\Re(C^{zz}_R(t)))$.}
\label{plot3}
\end{figure}

The spin-spin correlation function $C^{zz}_R(t)$ is plotted on Fig.~\ref{plot3} for both theoretical approaches. 
Before discussing the space-time correlation patterns, the Jordan-Wigner approximation implying $\mathcal{G}_k(t) = \langle \hat{c}^{\dag}_k \hat{c}_k \rangle_t \ll 2s$ and meaning that the quantum spin chain is well confined into the $z$ polarized phase has been verified. Concerning the correlation patterns, one can clearly notice a qualitative agreement. Indeed, the order of magnitude of the spin-spin correlations is very similar between the two plots. This means that both theoretical approaches provide qualitatively the same results and thus are consistent for the transverse Ising chain relatively deep into the $z$ polarized phase. 

\subsection{Quench dynamics for negative $\gamma$}\label{Sec:QuenchNegativeGamma}

In the following, we investigate the sudden global quench dynamics of the TFIM in the one-dimensional case from $\gamma = 0$ to $\gamma < 0$. As discussed in Sec.~\ref{sec:model} this situation can arise for example for a different choice of jump operators, associated to a projection on the spin-down rather than a spin-up component.
As we have already written, this situation tends to favor the quantum state $\ket{\uparrow}$ and we expect that the divergences discussed in the previous sections should disappear.

\begin{figure}[t]
\includegraphics[scale = 0.30]{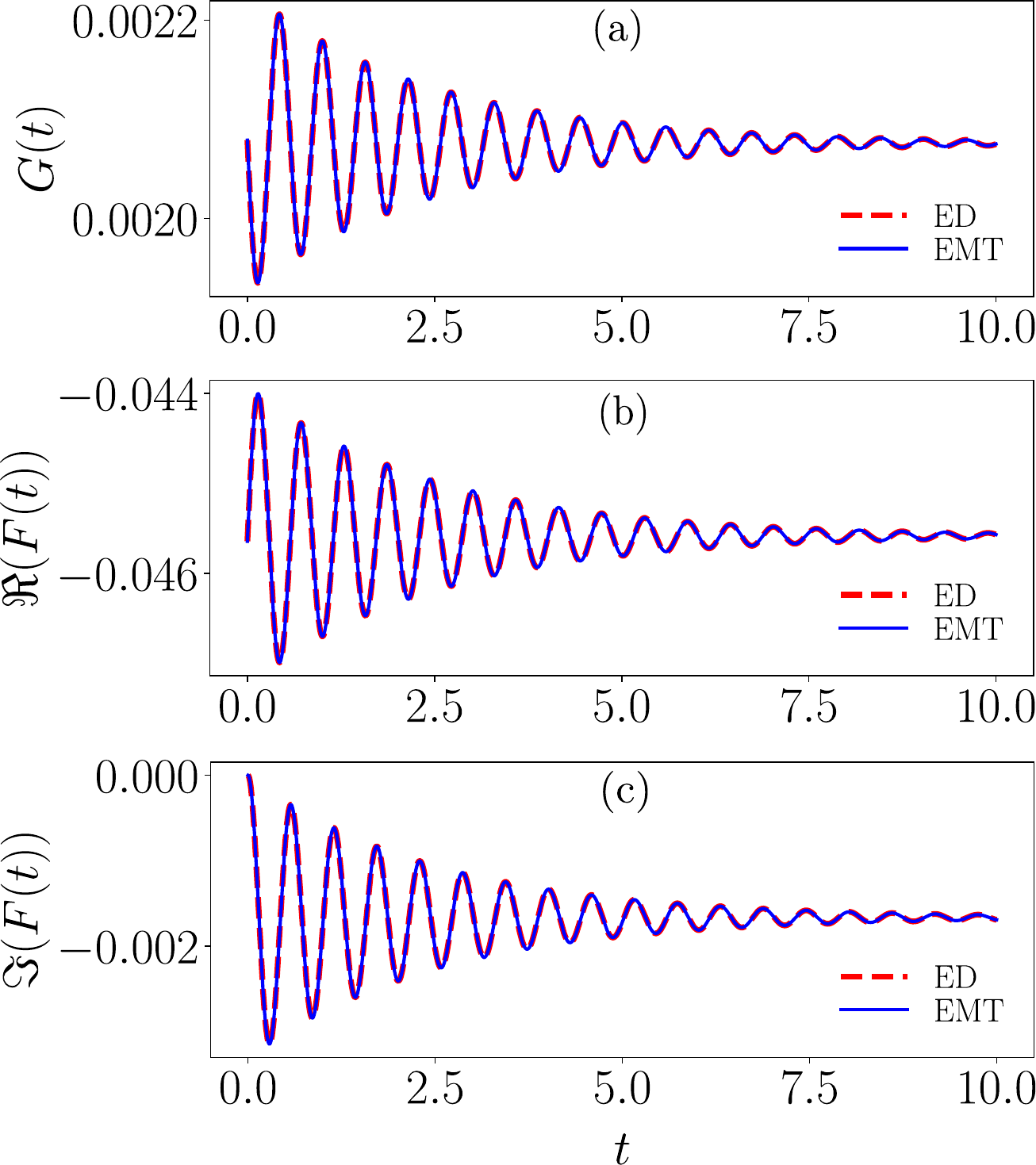}
\caption{Single-mode two-body correlators as a function of time $t$ where a sudden global quench has been considered on $\gamma$ from $\gamma = 0$ to $\gamma < 0$ for the non-Hermitian  transverse Ising chain confined in the $z$ polarized phase. (a) $G(t) = \langle \hat{a}^{\dag} \hat{a} \rangle_t$, (b) $\Re(F(t)) = \Re(\langle \hat{a}\hat{a} \rangle_t)$, (c) $\Im(F(t)) = \Im(\langle \hat{a}\hat{a} \rangle_t)$. For each plot, we compare the results found using exact diagonalization (see dashed red lines) and using the equation of motion technique (see solid blue lines). Numerical parameters : $h=5$, $\gamma = -0.2$, $J=1$, $N_{\mathrm{max}}=10$.}
\label{plot_single_mode_gamma_negative}
\end{figure}

We begin our study by performing the time-evolution with the single-mode Hamiltonian in Eq.~\eqref{H_single_mode}, where our results for $\gamma < 0$ are presented in Fig.~\ref{plot_single_mode_gamma_negative}. We observe a perfect agreement between the results found using the EoM and the ED. As expected, the dynamics is stabilized and no divergence occurs anymore. Note that the values of $G(t)$ are very small, which means that the system is close to the $\ket{\uparrow}$ state. 

\begin{figure}[b]
\includegraphics[scale = 0.45]{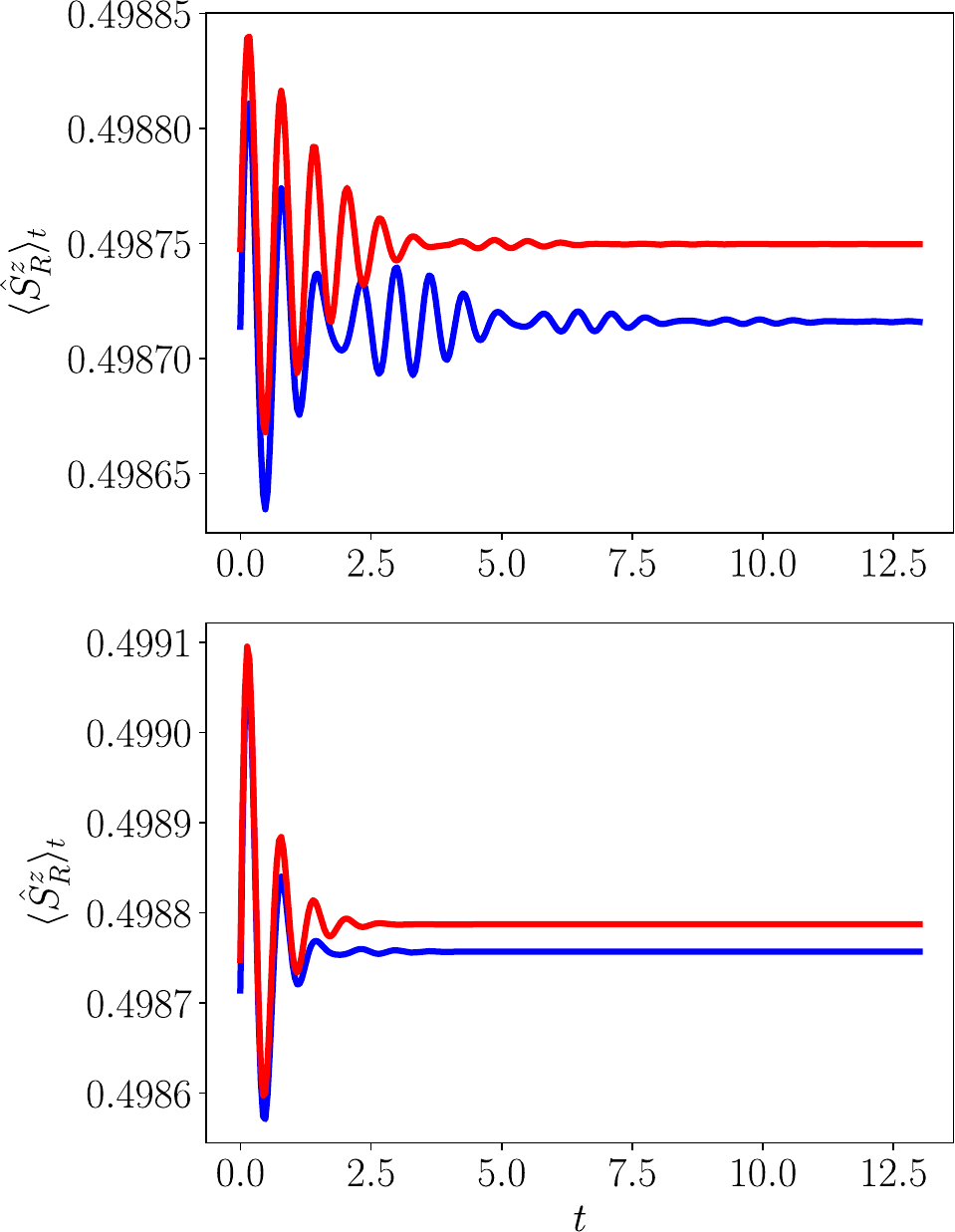}
\caption{Local magnetization $\langle \hat{S}^z_R\rangle_t$ for a sudden global quench on $\gamma$ from $\gamma = 0$ (Hermitian case) to $\gamma < 0$ (non-Hermitian case) for the transverse Ising chain in the $z$ polarized phase. For the profile associated to the bosonic approach, see solid blue line and Eq.~\eqref{szrt_boson} and for the one associated to the fermionic approach,
see solid red line and Eq.~\eqref{szrt_fermion}. The parameters of the simulation are: $N_s = 200$, $h=5$, $J=1$, $t_{\mathrm{i}}=0$, $t_{\mathrm{f}} = 13$, $\mathrm{steps} = 300$. 
(Top) $\gamma = -0.2$, (bottom) $\gamma = -0.9$.}
\label{plot_lm_gamma_negative}
\end{figure}

We now turn to the multi-mode problem where the time-evolution of the local magnetization for the 1D spin chain is investigated. 
In Fig.~\ref{plot_lm_gamma_negative} we present our results for the local magnetization computed both with the bosonic and the fermionic approaches;
we can observe that they are in qualitative agreement. 
In both cases, the local magnetization has values close but slightly below $1/2$ as  time increases because we start and remain close to an almost fully-polarized product state $\ket{\Psi_0} \simeq \ket{\uparrow \uparrow ... \uparrow \uparrow}$. 
The dynamics occurs at small times and then stabilizes to a final value larger than the initial one.

Using the fermionic approach, the stationary or final value of the local magnetization is given by the EoMs at Eq.~\eqref{fermion_diff_eq_fkt} and \eqref{fermion_diff_eq_gkt} by fixing each derivative to zero, i.e.~ $\mathrm{d}\mathcal{F}_k(t)/\mathrm{d}t = 0$ and $\mathrm{d}\mathcal{G}_k(t)/\mathrm{d}t = 0$. Then, the system to solve reads:

\begin{align}
 -8\Im(\mathcal{A}_k)\mathcal{F}_k\mathcal{G}_k -4i\mathcal{A}_k \mathcal{F}_k + 2i\mathcal{B}_k (1-2\mathcal{G}_k) & = 0, \\
 4\Im(\mathcal{A}_k)(|\mathcal{F}_k|^2 + \mathcal{G}_k - \mathcal{G}_k^2) + 4i\mathcal{B}_k\Re(\mathcal{F}_k) & = 0,
\end{align}

\noindent
and it yields the following equation for the correlator $\mathcal{G}_k$: 

\begin{align}
& 16\mathcal{B}_k^2 \Im(\mathcal{A}_k) + 64\Im(\mathcal{A}_k)(|\mathcal{A}_k|^2 - \mathcal{B}_k^2)\mathcal{G}_k + 512\Im(\mathcal{A}_k)^3\mathcal{G}_k^3 \nonumber \\
& + (64\Im(\mathcal{A}_k)(\mathcal{B}_k^2- |\mathcal{A}_k|^2) -256\Im(\mathcal{A}_k)^3 )\mathcal{G}_k^2  - 256\Im(\mathcal{A}_k)^3 \mathcal{G}_k^4 = 0.
\end{align}


\noindent
When solving the previous equation for each momentum $k$ and by taking the real and lowest solution for $\mathcal{G}_k$, the stationary value of the local magnetization $\langle \hat{S}^z_R \rangle_{\mathrm{f}}$ (the index '$\mathrm{f}$' stands for 'fermionic') reads as:

\begin{equation}
\langle \hat{S}^z_R \rangle_{\mathrm{f}} = \frac{1}{2} - \frac{1}{N_s} \sum_k \mathcal{G}_k.
\end{equation}

\noindent
Using the latter, we recover the value of the stationary local magnetization on Fig.~\ref{plot_lm_gamma_negative} when the same sets of parameters are considered. To extract the stationary local magnetization when using the fermionic approach, see the value of the local magnetization at large times on solid red lines.

Similarly, using the bosonic approach, the stationary value of the local magnetization is given by the EoMs at Eq.~\eqref{diff_eq_F_k_t} and \eqref{diff_eq_G} by fixing each derivative to zero, 
i.e.~ $\mathrm{d}F_k(t)/\mathrm{d}t = 0$ and $\mathrm{d}G_k(t)/\mathrm{d}t = 0$. The system to solve reads:

\begin{align}
4 \Im(A_k) F_k G_k -2iA_k F_k -iB_k(1+2G_k) &= 0, \\
-2B_k\Im(F_k)+2\Im(A_k)(|F_k|^2 + G_k + G_k^2) &= 0,
\end{align}

\noindent
which yields for $G_k$ the following equation:

\begin{align}
& -2B_k^2 \Im(A_k) + 8\Im(A_k)(|A_k|^2 - B_k^2)G_k + 64 \Im(A_k)^3 G_k^3 \nonumber \\
& + 8\Im(A_k)(|A_k|^2 - B_k^2 + 4\Im(A_k)^2)G_k^2 + 32\Im(A_k)^3G_k^4 = 0.
\end{align}

\noindent
When solving the previous equation for each momentum $k$ and by taking the real and lowest positive solution for $\mathcal{G}_k$, the stationary value of the local magnetization $\langle \hat{S}^z_R \rangle_{\mathrm{b}}$ (the index '$\mathrm{b}$' stands for 'bosonic') reads as:

\begin{equation}
\langle \hat{S}^z_R \rangle_{\mathrm{b}} = \frac{1}{2} - \frac{1}{N_s} \sum_k G_k.
\end{equation}

\noindent
Using the latter, we recover the value of the stationary local magnetization on Fig.~\ref{plot_lm_gamma_negative} when the same sets of parameters are considered. To extract the stationary local magnetization when using the bosonic approach, see the value of the local magnetization at large times on solid blue lines.

\begin{figure}[t]
\includegraphics[scale = 0.46]{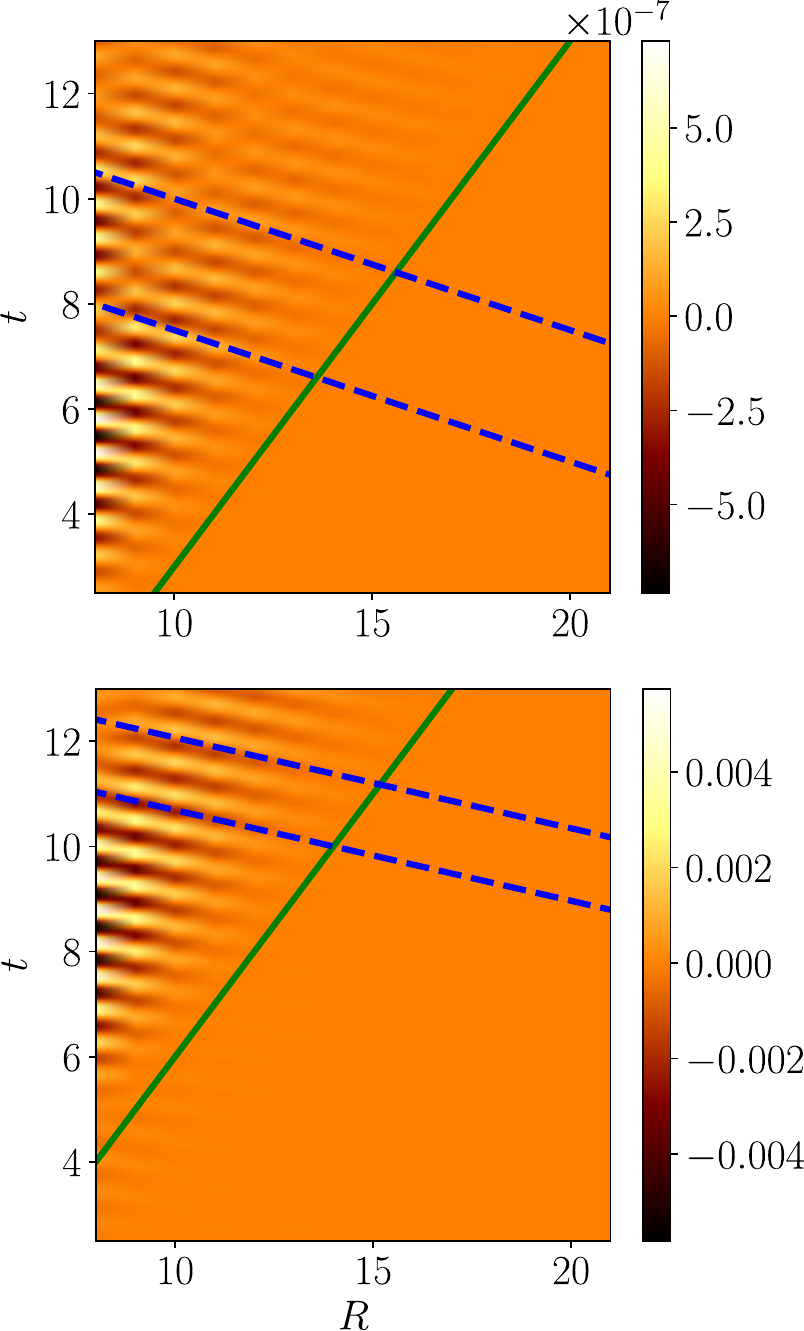}
\caption{Space-time plot of the one-body correlation function $G_R(t)$ for a sudden global quench on $\gamma$ of the TFIM in the paramagnetic phase. 
We start from the ground state at $\gamma = 0$ (Hermitian case) and evolve it in time with $\gamma < 0$ (non-Hermitian case) for the transverse Ising chain in the $z$ polarized phase. 
(Top) The results are obtained by integrating the EoM in Eq.~\eqref{diff_eq_G_k_t}. (Bottom) Theoretical guess of the correlation function characterised by Eq.~\eqref{G_R_t_th_guess}.
The parameters of the simulation are: $N_s = 200$, $h=5$, $\gamma = -0.2$, $J=1$.
The slope of the superimposed lines are: $V_{\mathrm{m}} = -4$ and $V_{\mathrm{CE}} = 1$ (top panel) and $V_{\mathrm{m}} = -5.8$ and $V_{\mathrm{CE}} = 1$ (bottom panel).}
\label{plot_correlation_pattern}
\end{figure}

We now focus on the spatial propagation of the quantum correlations induced by the dissipative quench for $\gamma < 0$ by analyzing the bosonic one-body correlation function defined at 
Eq.~\eqref{G_R_t_G_k_t}. We remind that the latter approximates the $\langle \hat{S}^{-}_R \hat{S}^{+}_0 \rangle_t $ spin correlator. Fig.~\ref{plot_correlation_pattern} presents the results of our numerical simulations, where we clearly notice the emergence of a light-cone structure in the space-time plot of $G_R(t)$. A clear CE propagating ballistically with a positive velocity $V_{\mathrm{CE}}$ is visible and highlighted by a green solid line in Fig.~\ref{plot_correlation_pattern}. Besides, in the vicinity of the CE the correlation function $G_R(t)$ displays a series of local minima and maxima which also spread ballistically at the negative velocity $V_{\mathrm{m}}$, highlighted in Fig.~\ref{plot_correlation_pattern} by several blue dashed lines. This behavior of the space-time correlations is reminiscent of the one found previously at Fig.~\ref{plot2} for a dissipative quench with $\gamma > 0$.

\section{Conclusion}
\label{sec:conclusion}

In this work we have discussed the linear spin-wave theory for the non-Hermitian TFIM in dimension $d=1,2,3$. Specifically we have first introduced the Holstein-Primakoff transformation and then performed a linear approximation to reduce the TFIM to a gas of non-Hermitian bosonic modes, diagonalizable with a complex bosonic Bogolyubov transformation. We have shown that this approach works well for the spectrum of the non-Hermitian quasiparticles. For the one-dimensional case, the quasiparticle dispersion relation found using the bosonic approach is in very good agreement with the exact one, that can be computed using a fermionic formalism based on the Jordan-Wigner transformation. This motivates the use of our theory for higher-dimensional lattices where exact solutions are not known.\\

We have then applied the linear spin wave theory to describe the non-Hermitian dynamics after a quench of the dissipative (non-Hermitian) coupling at time $t=0$. We have shown that for certain initial states and values of the dissipation the linear spin-wave approximation breaks down. While one could have anticipated an amplification of quasiparticle occupation numbers due to the non-Hermitian dynamics, leading to a violation of the harmonic approximation, we have shown that the breakdown is more severe and takes the form of a genuine finite-time divergence that we have explained using a simple analytical model.



We used the developed theory at short times to study the dissipative light-cone structure of quantum correlations induced by the non-Hermitian quench both in the one-dimensional and two-dimensional cases.\\ This work opens a number of questions about the possibility of applying spin-wave theory to study the dynamics of open quantum systems. In this simple non-Hermitian context we have shown that a naive application of spin-wave theory is limited to short times when $\gamma > 0$. Finding techniques for extending the results to longer times is an intriguing open project.


\acknowledgments
This research was supported by LabEx PALM (ANR-10-LABX-0039-PALM)
in Orsay and by Region Ile-de-France in the framework of DIM QuanTip.

\appendix

\section{Lindblad jump operators and the no-click limit}
\label{lindblad_appendix}
We give here further details concerning the derivation of the non-Hermitian Hamiltonian starting from a Lindblad master equation in the so called no-click limit.\\
The Lindblad master equation is given by :

\begin{equation}
\frac{\mathrm{d}}{\mathrm{d}t}\hat{\rho}(t) = -i \left[\hat{H}_{\mathrm{h}},\hat{\rho}(t)\right] + \sum_R \hat{L}_R \hat{\rho}(t) \hat{L}^{\dag}_R - \frac{1}{2} \left \{ \hat{L}^{\dag}_R \hat{L}_R, \hat{\rho}(t) \right \}, 
\label{lind}
\end{equation}

\noindent
where $\hat{H}_{\mathrm{h}}$ refers to the Hamiltonian of the spin-$1/2$ (Hermitian) transverse-field Ising chain, see Eq.~\eqref{H_ising_chain} with $\gamma = 0$. The latter can be rewritten in terms of a non-Hermitian Hamiltonian $\hat{H}$: 

\begin{equation}
\frac{\mathrm{d}}{\mathrm{d}t}\hat{\rho}(t) = -i \left( \hat{H}\hat{\rho}(t) - \hat{\rho}(t)\hat{H}^{\dag} \right) + \sum_R \hat{L}_R \hat{\rho}(t) \hat{L}^{\dag}_R,
\end{equation}

\noindent
where $\hat{H}$ reads:

\begin{equation}
\hat{H} = \hat{H}_{\mathrm{h}} - \frac{i}{2} \sum_R \hat{L}^{\dag}_R \hat{L}_R. 
\label{nh}
\end{equation}

\noindent
Considering $\hat{L}_R = \sqrt{2\gamma}(1/2 + \hat{S}^z_R)$ for $\gamma > 0$, the non-Hermitian Hamiltonian
$\hat{H}$ is given by:

\begin{equation}
\hat{H} = \hat{H}_{\mathrm{h}} - \frac{i}{2} \sum_R \hat{L}^{\dag}_R \hat{L}_R = \hat{H}_{\mathrm{h}} -i\gamma \sum_R \hat{S}^z_R, 
\end{equation}

\noindent
where a pure imaginary constant term has been disregarded. As expected, we recover the expression at Eq.~\eqref{H_ising_chain}
for the non-Hermitian Hamiltonian $\hat{H}$. The reasoning is similar for $\gamma < 0$ where the Lindblad jump operator is given by $\hat{L}_R = \sqrt{2|\gamma|}(1/2 - \hat{S}^z_R)$. \\

\noindent

The Lindblad master equation can be unravelled in a set of quantum trajectories, describing the evolution of the system conditioned to a set of measurement outcomes. While different unravelling schemes exist, a relevant one to understand the emergence of the non-Hermitian Hamiltonian is the so-called quantum jump protocol~\cite{daley2014}. Here the evolution of the system is described by a stochastic Schr\"odinger equation where at random times a jump operator $\hat{L}_{R}$ is applied to the state while in between quantum jumps the evolution of the system is non-unitary and driven by the non-Hermitian Hamiltonian in Eq.~(\ref{nh}), which includes the measurement back-action. From this perspective, the non-Hermitian dynamics corresponds to a post-selection of the so called no-click quantum trajectory, where all the measurement outcomes correspond to a non-click event. While this trajectory, as any other characterized by a well defined measurement record, is exponentially rare, it is important to emphasize that the non-Hermitian Hamiltonian controls also the evolution in between quantum jumps. As such its study can also provide insights on the dynamics of typical quantum trajectories~\cite{ashida2020}.
%
%
%
%
%
%

\section{Bosonic and fermionic Bogolyubov transformations}
\label{theta_k_appendix}

\begin{figure}[t]
\centering
\includegraphics[width=\columnwidth]{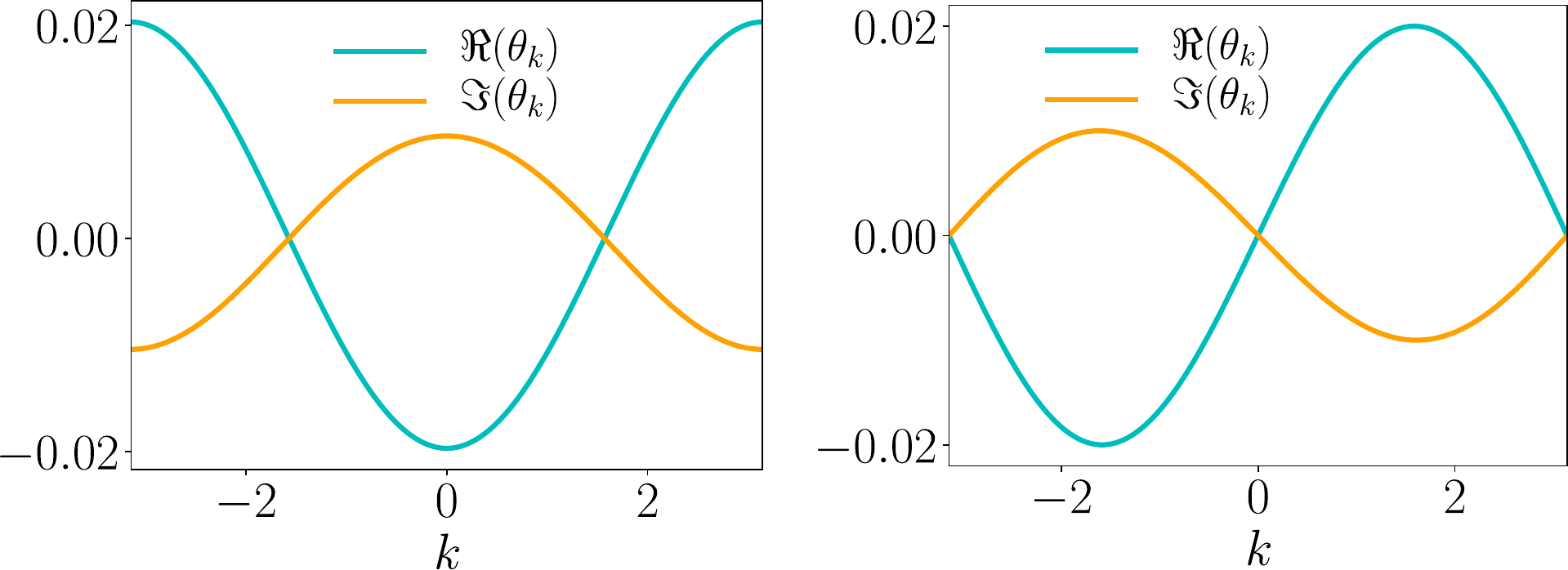}
\caption{Momentum-dependent function $\theta_k$ where $\Re(\theta_k)$ (solid cyan line) and $\Im(\theta_k)$ (solid orange line) are plotted for the complex bosonic Bogolyubov transformation
(left) and the complex fermionic Bogolyubov transformation (right). The numerical parameters are : $J=1$, $h=20$ and $\gamma=10$.}
\label{thetak}
\end{figure}

Concerning the complex bosonic Bogolyubov transformation for the one-dimensional case, the momentum-dependent function $\theta_k$ has the following expression
\begin{equation}
\theta_k = \mathrm{arctanh}\left(-B_k/A_k\right),
\end{equation}
where $A_k = h + (J/2)\cos(k) + i\gamma$ and $B_k = (J/2)\cos(k)$, see Fig.~\ref{thetak} for an example. For the complex fermionic Bogolyubov transformation, $\theta_k$ is given by 
\begin{equation}
\theta_k = \mathrm{arctan}\left(i \mathcal{B}_k / \mathcal{A}_k \right),
\end{equation}
where $\mathcal{A}_k = (J/4)\cos(k) + (h+i\gamma)/2$ and $\mathcal{B}_k = -i(J/4)\sin(k)$, see Fig.~\ref{thetak} for an example.

\section{Fourier transform of the bosonic and fermionic operators}
\label{fourier_transform_op}

The convention of the Fourier transform is the following for the bosonic and fermionic operators respectively for a $D$-dimensional hypercubic lattice

\begin{align}
& \hat{a}^{\dag}_{\mathbf{R}} = \frac{1}{N_s^{D/2}}\sum_{\mathbf{k}} e^{-i\mathbf{k}\mathbf{R}}\hat{a}^{\dag}_{\mathbf{k}},~~~ \hat{a}_{\mathbf{R}} = \frac{1}{N_s^{D/2}}\sum_{\mathbf{k}} e^{i\mathbf{k}\mathbf{R}}\hat{a}_{\mathbf{k}}, \\
& \hat{c}^{\dag}_{\mathbf{R}} = \frac{1}{N_s^{D/2}}\sum_{\mathbf{k}} e^{-i\mathbf{k}\mathbf{R}}\hat{c}^{\dag}_{\mathbf{k}},~~~ \hat{c}_{\mathbf{R}} = \frac{1}{N_s^{D/2}}\sum_{\mathbf{k}} e^{i\mathbf{k}\mathbf{R}}\hat{c}_{\mathbf{k}}. 
\end{align}

%

\section{Equations of motion - bosonic approach}
\label{EMT}

\begin{figure}[t]
\centering
\includegraphics[scale = 0.30]{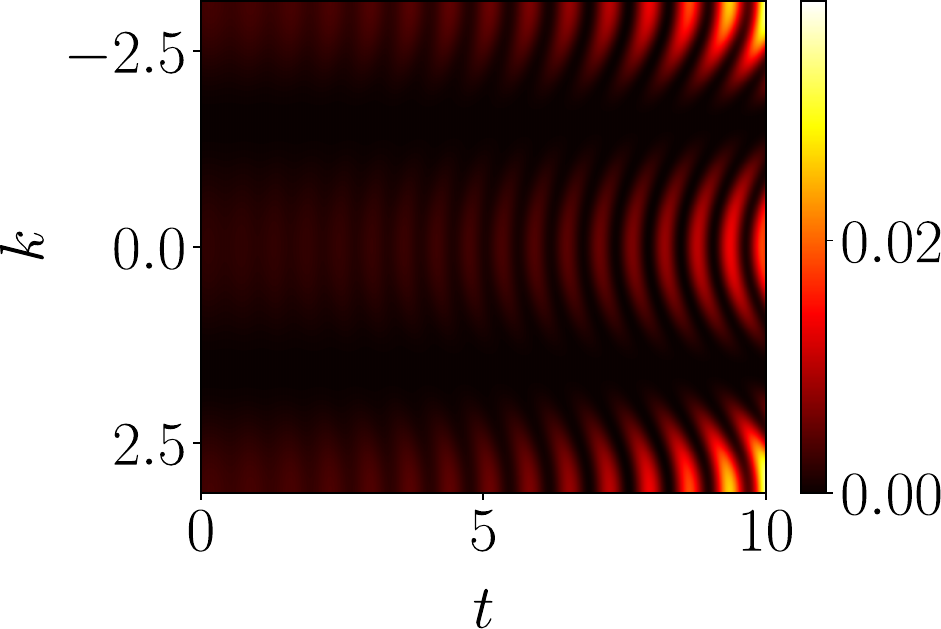}
\caption{Numerical momentum-time correlator $G_k(t) = \langle \hat{a}^{\dag}_k \hat{a}_k \rangle_t$ for a sudden global quench on $\gamma$ from $\gamma = 0$ (Hermitian case) to $\gamma \neq 0$ (non-Hermitian case) for the transverse Ising chain in the $z$ polarized phase. The numerical parameters are : $N_s = 1000$, $h=5$, $\gamma = 0.2$, $J=1$.}
\label{momentum_correlators}
\end{figure}

Starting from the time-evolved quantum state $\ket{\Psi(t)}$ given by

\begin{equation}
\ket{\Psi(t)} = \frac{e^{-i\hat{H}t}\ket{\Psi_0}}{||e^{-i\hat{H}t}\ket{\Psi_0}||},
\end{equation}

\noindent
we calculate its derivative with respect to the time $t$ which can be written as  

\begin{align}
& \frac{\mathrm{d}}{\mathrm{d}t} \ket{\Psi(t)} = -i \hat{H} \ket{\Psi(t)} - \frac{i}{2} \langle \Psi(t)| \hat{H}^{\dag}- \hat{H}|\Psi(t) \rangle \ket{\Psi(t)}.
\end{align}

\noindent
Using the latter and the equation below 

\begin{align}
& \frac{\mathrm{d}}{\mathrm{d}t} \bra{\Psi(t)} = \frac{\mathrm{d}}{\mathrm{d}t} \ket{\Psi(t)}^{\dag} = 
\left( \frac{\mathrm{d}}{\mathrm{d}t} \ket{\Psi(t)} \right)^{\dag},
\end{align}

\noindent
we can deduce the equation of motion for the time-dependent expectation value of the observable $\hat{A}$ denoted by $G_A(t) = \langle \Psi(t)|\hat{A}|\Psi(t)\rangle = \langle \hat{A} \rangle_t$ and given by

\begin{align}
& \frac{\mathrm{d}}{\mathrm{d}t} G_A(t) = i \langle \hat{H}^{\dag}\hat{A}-\hat{A}\hat{H}\rangle_t + i\langle \hat{H}-\hat{H}^{\dag}\rangle_t G_A(t).
\end{align}

\noindent
The Hamiltonian $\hat{H}$ used to perform the time evolution is given at Eq.~\eqref{H}. After some analytical calculations where the bosonic Wick theorem together with the momentum conservation
are used, we find the coupled complex differential equations at Eq.~\eqref{diff_eq_F_k_t} and \eqref{diff_eq_G} for the correlator $F_k(t) = \langle \hat{a}_k \hat{a}_{-k} \rangle_t$
and $G_k(t) = \langle \hat{a}^{\dag}_k \hat{a}_k \rangle_t$ respectively. The initial conditions $F_k(0) = \langle \hat{a}_{-k} \hat{a}_k \rangle_0$ and 
$G_k(0) = \langle \hat{a}^{\dag}_k \hat{a}_k \rangle_0$, with $\langle ...\rangle_0 = \langle \Psi_0 | ... | \Psi_0 \rangle$ and $\ket{\Psi_0} = \ket{\Psi(0)} 
= \ket{\mathrm{GS}(\hat{H}_i)}$ where the initial (pre-quench) Hermitian Hamiltonian $\hat{H}_i$ has the following expression: 

\begin{align}
& \hat{H}_i = \frac{1}{2} \sum_k A_{k,i} \left( \hat{a}^{\dag}_{k,i} \hat{a}_{k,i} + \hat{a}_{-k,i} \hat{a}_{-k,i}^{\dag} \right) \nonumber \\
& ~~~~~~~ + B_k \left( \hat{a}^{\dag}_{k,i} \hat{a}^{\dag}_{-k,i} + \hat{a}_{k,i} \hat{a}_{-k,i} \right),
\end{align}

\noindent
where $A_{k,i} = h + (J/2)\cos(k)$ and $B_k = (J/2)\cos(k)$, can be deduced by following the procedure below. In order to calculate the correlators at time $t=0$,
we rely on the following bosonic Bogolyubov transformation  

\begin{align}
& \hat{a}_{k,i} = u_{k,i} \hat{\beta}_{k,i} + v_{-k,i} \hat{\beta}^{\dag}_{-k,i}, \\
& \hat{a}_{k,i}^{\dag} = u_{k,i}\hat{\beta}^{\dag}_{k,i} + v_{-k,i}\hat{\beta}_{-k,i},
\end{align}

\noindent
where the real momentum-dependent functions $u_{k,i}$ and $v_{-k,i}$ are parametrized by $\alpha_{k,i}$ and given by  

\begin{align}
& u_{k,i} = \cosh(\alpha_{k,i}), \\
& v_{-k,i} = \sinh(\alpha_{k,i}),
\end{align}

\noindent
where $\alpha_{k,i}$ is defined as 

\begin{equation}
\alpha_{k,i} = \frac{1}{2}\mathrm{arctanh} \left(-\frac{B_k}{A_{k,i}}\right).
\end{equation}

\noindent
By replacing the bosonic annihilation and creation operators $\hat{a}_k$ and $\hat{a}^{\dag}_k$ by $\hat{a}_{k,i}$ and
$\hat{a}^{\dag}_{k,i}$ respectively in the correlators and by injecting the expression of $\hat{a}_{k,i}$ and $\hat{a}^{\dag}_{k,i}$ 
in terms of the bosonic Bogolyubov operators denoted by $\hat{\beta}_{k,i}$ and $\hat{\beta}^{\dag}_{k,i}$, we find: 

\begin{align}
& F_k(0) = u_{k,i}v_{k,i}, \\
& G_k(0) = v_{k,i}^2. 
\end{align}

\section{Theoretical guess for the one-body correlation function $G_R(t)$}
\label{th_guess}

In order to unveil a theoretical expression for $G_R(t)$, it requires to make an analytical guess for $G_k(t)$ according to Eq.~\eqref{G_R_t_G_k_t}.
From Fig.~\ref{momentum_correlators} where the momentum-time behavior of $G_k(t)$ is displayed, we can make the following theoretical guess: 
 
\begin{align}
& G_k(t) = F_k \cos(2\Re(E_k)t)\exp(2\Im(E_k)t), 
\end{align}

\noindent
where the momentum-dependent $F_k$ is defined as 

\begin{align}
& F_k = \exp \left(-\frac{(k-\pi)^2}{2\sigma^2} \right) + \exp \left(-\frac{(k+\pi)^2}{2\sigma^2} \right) \nonumber \\
& ~~~~~~ + \exp \left(-\frac{k^2}{2\sigma^2} \right).
\end{align}
  
\begin{figure}[t]
\includegraphics[scale = 0.29]{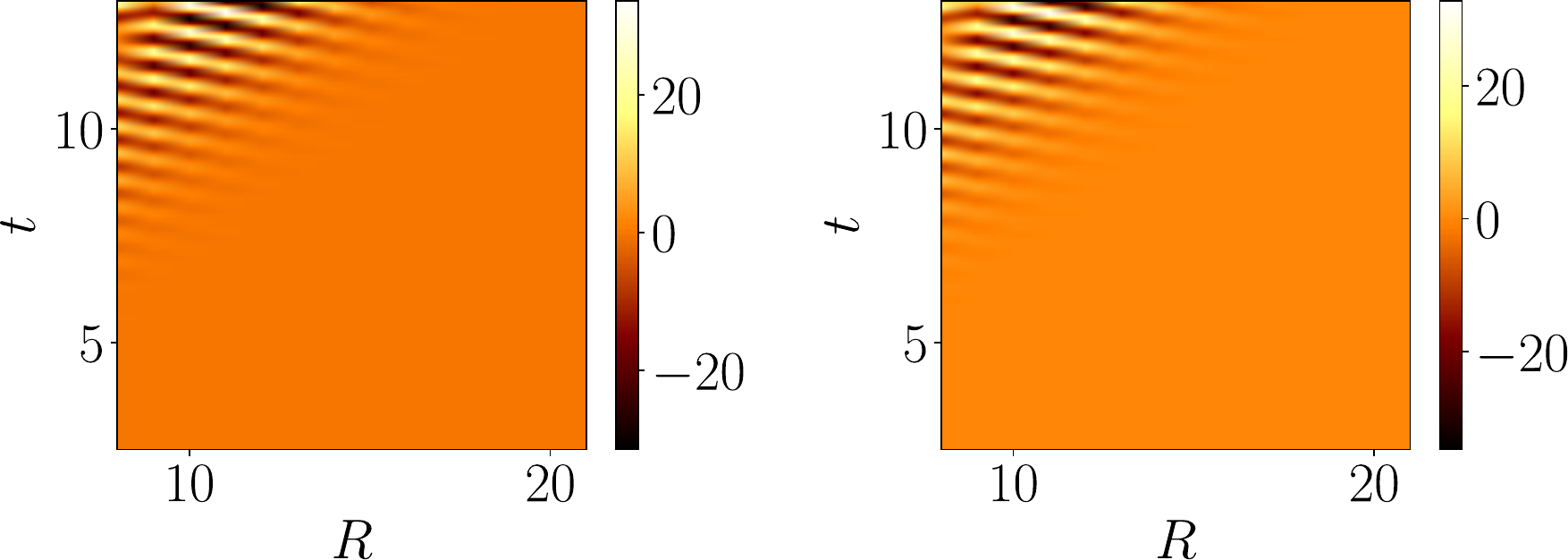}
\caption{Space-time one-body correlation function $G_R(t)$, see Eq.~\eqref{G_R_t_G_k_t}, for a sudden global quench on $\gamma$ from $\gamma = 0$ (Hermitian case) 
to $\gamma \neq 0$ (non-Hermitian case) for the transverse Ising chain in the $z$ polarized phase. Plot of the theoretical guess at Eq.~\eqref{grt}(left) 
and the theoretical guess with the approximation $E_k \approx A_k$ (right), see Eq.~\eqref{grt_bis}. The parameters are : $N_s = 200$, $h=5$, $\gamma = 0.2$, $J=1$, $t_{\mathrm{i}}=0$, 
$t_{\mathrm{zoom}} = 2.5$, $t_{\mathrm{f}} = 13$, $\mathrm{steps} = 300$, $R_{\mathrm{i}} = 8$, $R_{\mathrm{f}} = 20$, $\sigma = 1$.}
\label{tls}
\end{figure}

\noindent 
Note that the parameter $\sigma$ is deduced directly and qualitatively from the numerical result of $G_k(t)$ on Fig.~\ref{momentum_correlators} by using the 
relation $\mathrm{FWHM} \simeq 2.35\sigma$. We obtain $\sigma = 1$. As a consequence, the theoretical guess of $G_R(t)$ is given by  
 
\begin{align}
& G_R(t) = \frac{1}{N_s} \sum_k F_k \cos(kR) \cos(2\Re(E_k)t)\exp(2\Im(E_k)t), \label{grt}
\end{align}

\noindent
where $E_k$ denotes the final (post-quench) excitation spectrum corresponding to the quasiparticle dispersion relation of the non-Hermitian transverse Ising chain, see Eq.~\eqref{spectrum}. 
In order to simplify Eq.~\eqref{grt}, we consider the approximation $E_k \approx A_k$ valid for $h \gg J$ which leads to the final expression: 

\begin{align}
& G_R(t) = \frac{1}{N_s} \sum_k F_k \cos(kR) \cos(2\Re(A_k)t)\exp(2\Im(A_k)t),
\label{grt_bis}
\end{align}

\noindent
where the momentum-dependent function $A_k$ is given by $A_k = h + (J/2)\cos(k) + i\gamma$. Note that the approximation $E_k \approx A_k$ works very well in the regime where 
$h \gtrsim J$ according to Fig.~\ref{tls} where the ratio $h/J = 5$ is considered and for which the two space-time patterns of $G_R(t)$ are identical.

\section{Numerical technique to track the correlation edge}
\label{ce_track_appendix}

\begin{figure}[t]
\includegraphics[scale = 0.28]{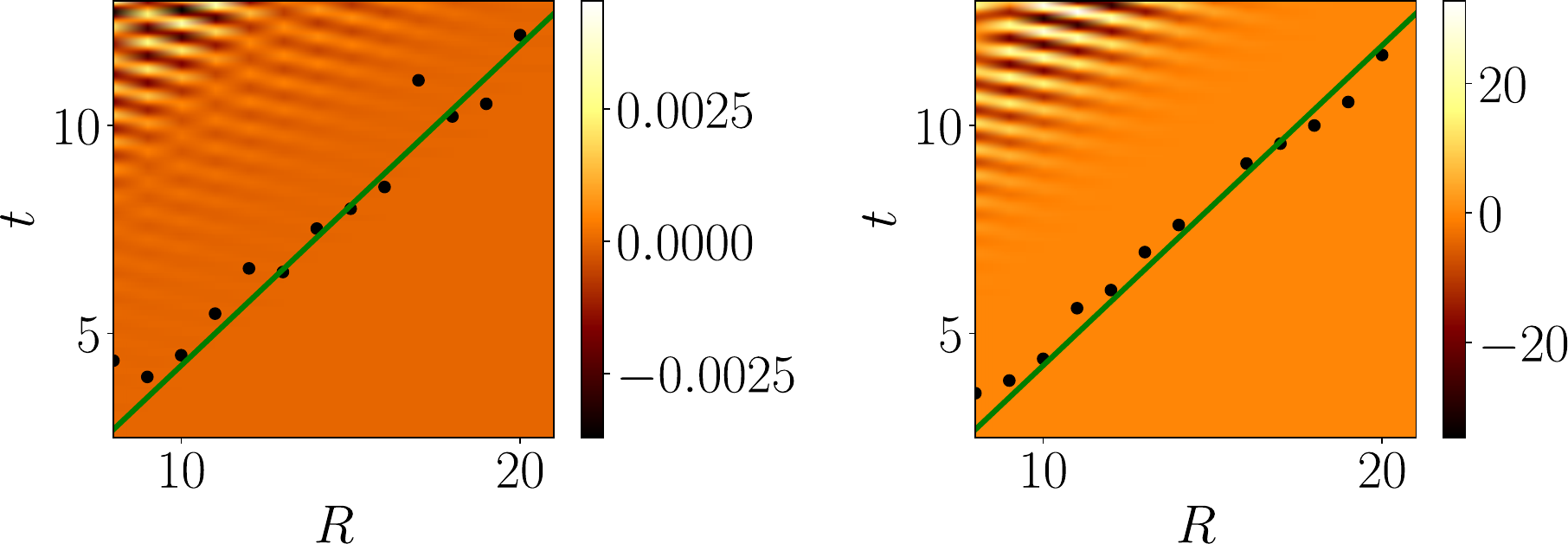}
\caption{Space-time one-body correlation function $G_R(t)$ for a sudden global quench on $\gamma$ from $\gamma = 0$ (Hermitian case) 
to $\gamma \neq 0$ (non-Hermitian case) for the transverse Ising chain in the $z$ polarized phase. Numerical result found using the equation of motion technique to deduce 
$G_k(t)$, see Eq.~\eqref{G_R_t_G_k_t} (left) and theoretical guess, see Eq.~\eqref{G_R_t_th_guess} (right). The parameters are : $N_s = 200$, $h=5$, $\gamma = 0.2$, $J=1$, $t_{\mathrm{i}}=0$, $t_{\mathrm{zoom}} = 2.5$, $t_{\mathrm{f}} = 13$,
$\mathrm{steps} = 300$, $R_{\mathrm{i}} = 8$, $R_{\mathrm{f}} = 20$, $\sigma = 1$. $V_{\mathrm{CE}} = 1.3$ is found for each plot.}
\label{track_ce}
\end{figure}

The correlation edge is found by tracking for each lattice $R$ the time $t_R$ also called activation time where the correlation reaches a very small percentage of the maximal value of the
correlations, see black dots on Fig.~\ref{track_ce}. The different points $(R,t_R)$ form a linear profile, see solid green line on Fig.~\ref{track_ce}, where the corresponding 
velocity $V_{\mathrm{CE}}$ is given by the slope.

\section{Equations of motion - fermionic approach}
\label{EMT_fermions}

We start from the equation of motion for the time-dependent expectation value of the observable $\hat{A}$ denoted by $G_A(t) = \langle \Psi(t)|\hat{A}|\Psi(t)\rangle = \langle 
\hat{A} \rangle_t$ and given by

\begin{align}
& \frac{\mathrm{d}}{\mathrm{d}t} G_A(t) = i \langle \hat{H}^{\dag}\hat{A}-\hat{A}\hat{H}\rangle_t + i\langle \hat{H}-\hat{H}^{\dag}\rangle_t G_A(t).
\end{align}

\noindent
The latter is derived in Appendix \ref{EMT}. The Hamiltonian $\hat{H}$ used to perform the time evolution is given at Eq.~\eqref{H_fermionic_quadratic}. After some analytical calculations where the fermionic Wick theorem together with the momentum conservation are used, we find the two coupled complex differential equations at Eq.~\eqref{fermion_diff_eq_fkt} and \eqref{fermion_diff_eq_gkt} for the correlator $\mathcal{F}_k(t) = \langle \hat{c}_k \hat{c}_{-k} \rangle_t$ and $\mathcal{G}_k(t) = \langle \hat{c}^{\dag}_k \hat{c}_k \rangle_t$ respectively. The initial conditions $\mathcal{F}_k(0) = \langle \hat{c}_{k} \hat{c}_{-k} \rangle_0$ and $\mathcal{G}_k(0) = \langle \hat{c}^{\dag}_k \hat{c}_k \rangle_0$, 
with $\langle ...\rangle_0 = \langle \Psi_0 | ... | \Psi_0 \rangle$ and $\ket{\Psi_0} = \ket{\Psi(0)} = \ket{\mathrm{GS}(\hat{H}_i)}$ where the initial (pre-quench) Hermitian
Hamiltonian $\hat{H}_i$ has the following expression  

\begin{align}
& \hat{H}_i = \sum_k \mathcal{A}_{k,i} \left(\hat{c}^{\dag}_{k,i} \hat{c}_{k,i} - \hat{c}_{-k,i}\hat{c}^{\dag}_{-k,i} \right) \nonumber \\
& ~~~~~~~ + \mathcal{B}_k \left(\hat{c}^{\dag}_{k,i} \hat{c}^{\dag}_{-k,i} + \hat{c}_{k,i} \hat{c}_{-k,i} \right),
\end{align}

\noindent
where $\mathcal{A}_{k,i} = (J/4)\cos(k) + h/2$ and $\mathcal{B}_k = -i(J/4)\sin(k)$, can be deduced by following the procedure below. In order to calculate the correlators at time $t=0$,
we rely on the following fermionic Bogolyubov transformation  

\begin{align}
& \hat{c}^{\dag}_{-k,i} = \cos\left(\frac{\theta_{k,i}}{2} \right) \hat{\eta}^{\dag}_{-k,i} + i \sin\left(\frac{\theta_{k,i}}{2} \right) \hat{\eta}_{k,i}, \\
& \hat{c}_{k,i} = \cos\left(\frac{\theta_{k,i}}{2} \right) \hat{\eta}_{k,i} + i \sin\left(\frac{\theta_{k,i}}{2} \right) \hat{\eta}^{\dag}_{-k,i},
\end{align}

\noindent
where the real momentum-dependent function $\theta_{k,i}$ is defined as 

\begin{equation}
\theta_{k,i} = \mathrm{arctan}\left(i \frac{\mathcal{B}_k}{\mathcal{A}_{k,i}} \right).
\end{equation}

\noindent
By replacing the fermionic annihilation and creation operators $\hat{c}_k$ and $\hat{c}^{\dag}_k$ by $\hat{c}_{k,i}$ and
$\hat{c}^{\dag}_{k,i}$ respectively in the correlators and by injecting the expression of $\hat{c}_{k,i}$ and $\hat{c}^{\dag}_{k,i}$ 
in terms of the fermionic Bogolyubov operators denoted by $\hat{\eta}_{k,i}$ and $\hat{\eta}^{\dag}_{k,i}$, we find 

\begin{align}
& \mathcal{F}_k(0) = -i \sin\left(\frac{\theta_{k,i}}{2} \right)\cos\left(\frac{\theta_{k,i}}{2} \right), \\
& \mathcal{G}_k(0) = \sin^2\left(\frac{\theta_{k,i}}{2} \right).
\end{align}

\section{Correlation edge velocity - square lattice}
\label{cev}

\begin{figure}[h]
\includegraphics[scale = 0.33]{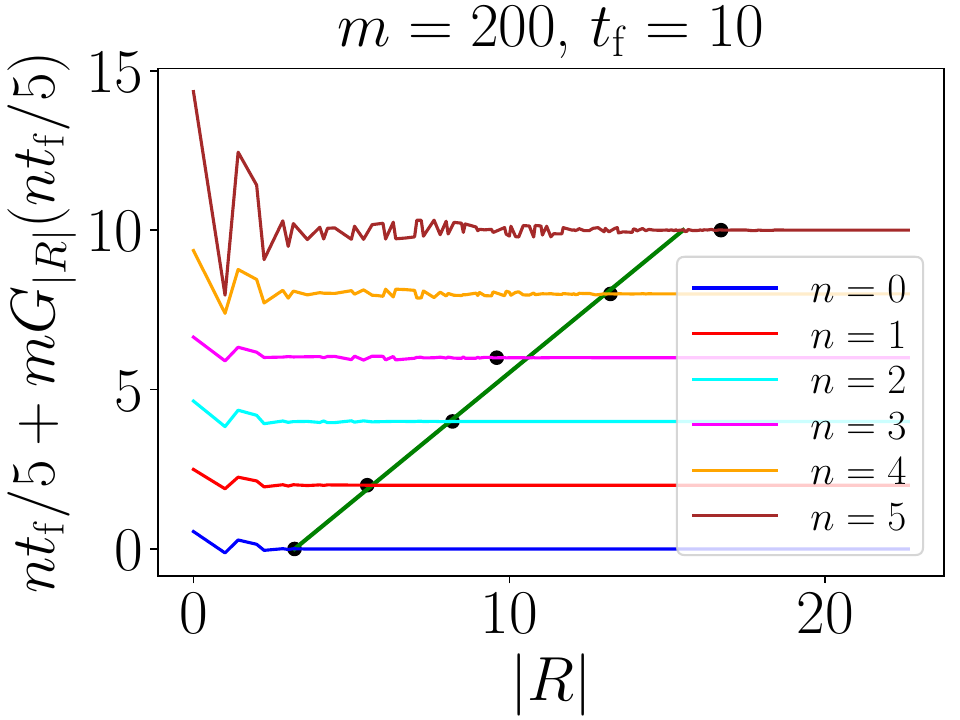}
\caption{One-body correlation function $G_{|R|}(t)$ for a sudden global quench on $\gamma$ from $\gamma = 0$ (Hermitian case) to $\gamma \neq 0$ (non-Hermitian case) for the transverse Ising model on a square lattice and confined in the $z$ polarized phase. $|R|$ represents the distance from the central site of the square lattice. The parameters are : $N_s = 100$, $h=5$, $\gamma = 0.2$, $J=1$, $t_{\mathrm{i}}=0$, $t_{\mathrm{f}} = 10$, $\mathrm{steps} = 300$.}
\label{corr_square_lattice_distance_from_center}
\end{figure}

We propose here another numerical technique in order to determine the correlation edge velocity $V_{\mathrm{CE}}$. The correlations are represented as a function of $|R|$ corresponding to 
the distance from the central site of the square lattice. The latter are reported on Fig.~\ref{corr_square_lattice_distance_from_center} for different times. For each value of the time, the 
distance $|R|^*$ is tracked such that above this value the correlations are zero corresponding to the black dots. These dots have a linear profile fitted by the solid green line where the 
slope corresponds to the solid green line. We find $V_{\mathrm{CE}} \simeq 1.2$. 

\section{Single-mode model with a nonlinearity}
\label{nl}

\begin{figure}[h!]
\centering
\includegraphics[scale = 0.41]{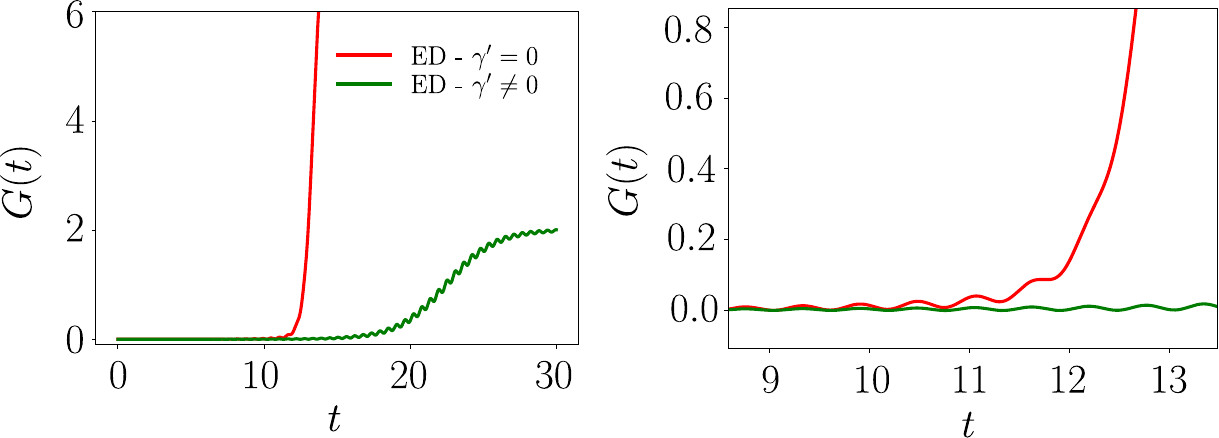} \\
\caption{Single-mode two-body correlator $G(t) = \langle \hat{a}^{\dag} \hat{a} \rangle_t$ as a function of time $t$ where a sudden global quench has been considered on the dissipative strength $\gamma$ from $\gamma = 0$ to $\gamma > 0$ for the non-Hermitian transverse Ising chain, with a quartic (interaction) term governed by the imaginary energy $-i\gamma'/2$, and confined in the $z$ polarized phase. (Left) The results found using ED for the non-interacting case $\gamma'=0$ (see red line) is compared to the interacting case where $\gamma' = 0.05$ (see green line).
(Right) Zoom of the single-mode two-body correlator $G(t) = \langle \hat{a}^{\dag} \hat{a} \rangle_t$. The numerical parameters are: $h = 5$, $\gamma = 0.2$, $J = 1$.}
\label{fig_nl}
\end{figure} 

We study here the effect of a weak nonlinearity on the finite-time divergence for the single-mode model. The latter is represented by a non-Hermitian quartic term
leading for the Hamiltonian $\hat{H}_s$ to: 

\begin{equation}
\hat{H}_s = i \frac{\gamma}{2} \hat{a}^{\dag}\hat{a} - i\frac{\gamma'}{4} \hat{a}^{\dag}\hat{a}\hat{a}^{\dag}\hat{a},
\end{equation}

\noindent
where $\gamma, \gamma'> 0$. The prefactor $1/2$ and $1/4$ are for convenience. The non-normalised time-evolved quantum state $\ket{\tilde{\Psi}(t)} = e^{-i\hat{H}_s t} \ket{\Psi(0)}$ now reads:

\begin{equation}
\ket{\tilde{\Psi}(t)} = \frac{1}{\sqrt{\cosh(r)}} \sum_{n=0}^{+\infty} \left(-e^{i\phi} \tanh(r) e^{(\gamma - \gamma'n)t}\right)^n \frac{\sqrt{(2n)!}}{2^n n!} \ket{2n}.
\end{equation}

\noindent
After some analytical calculations, its norm is given by:

\begin{equation}
||\ket{\tilde{\Psi}(t)} ||^2 = \frac{1}{\cosh(r)} \sum_{n=0}^{+\infty} a_n(t),
\end{equation}

\noindent
where the time-dependent series $a_n(t)$ is defined as:

\begin{equation}
a_n(t) = \left(\tanh(r)e^{(\gamma-\gamma'n)t} \right)^{2n} \frac{(2n)!}{2^{2n}(n!)^2}.
\end{equation}

\noindent
To know whether the series converges or not, we have to calculate the following quantity according to the D'Alembert criterion:

\begin{equation}
\lim\limits_{n \rightarrow +\infty} \frac{a_{n+1}(t)}{a_{n}(t)}.
\end{equation}

\noindent
After some analytical calculations, the ratio can be expressed as:

\begin{equation}
\frac{a_{n+1}(t)}{a_{n}(t)} = \tanh^2(r) e^{2\gamma t} e^{-2(2n+1)\gamma' t} \frac{(2n+2)(2n+1)}{4(n+1)^2}.
\end{equation}

\noindent
Consequently, this leads to: 

\begin{equation}
\lim\limits_{n \rightarrow +\infty} \frac{a_{n+1}(t)}{a_{n}(t)} = 0,
\end{equation}

\noindent
meaning that the series and thus the norm converges. To sum up, when a nonlinearity is considered and represented by a non-Hermitian interaction term, the finite-time divergence is cured. In what follows, we investigate numerically the single-mode model including the weak nonlinearity using ED. The single-mode Hamiltonian of the non-Hermitian transverse Ising chain with the interaction term is given by:

\begin{equation}
\hat{H} = \frac{h + (J/2) + i\gamma}{2}(\hat{a}^{\dag}\hat{a} + \hat{a}\hat{a}^{\dag}) + \frac{J}{4}(\hat{a}^{\dag}\hat{a}^{\dag} + \hat{a}\hat{a}) - \frac{i\gamma'}{2} \hat{a}^{\dag}\hat{a}\hat{a}^{\dag}\hat{a}.
\label{new_H}
\end{equation}

\noindent
From Fig.~\ref{fig_nl}, the finite-time divergence is only visible in the non-interacting case. This means that a weak nonlinearity is able to cure the divergence. To get an estimation of  $t_{\mathrm{f}}'$ the specific time for which the effect of the nonlinearity becomes relevant, we consider the following condition: 

\begin{equation}
t = t_{\mathrm{f}}':~\gamma \langle \hat{a}^{\dag}\hat{a} \rangle_t \approx \gamma' \langle \hat{a}^{\dag}\hat{a} \hat{a}^{\dag}\hat{a} \rangle_t,
\end{equation}

\noindent
where the time-evolved quantum state is computed from the Hamiltonian $\hat{H}$ of the single-mode model without the nonlinearity, see Eq.~\ref{new_H} with $\gamma'=0$.
For $\gamma = 0.2$ and $\gamma' = 0.05$, we find $t_{\mathrm{f}}' \simeq 11.2$ which is a good estimation according to Fig.~\ref{fig_nl}.

\noindent
We have demonstrated analytically and numerically that a non-Hermitian quartic term can cure the finite-time divergence. However, such term does not appear in the (multi-mode) Hamiltonian of the non-Hermitian short-range transverse Ising chain, see Eq.~\ref{H_ising_chain}. Indeed, the non-Hermitian part $\hat{H}_{\mathrm{nh}}$ is given by:  

\begin{equation}
 \hat{H}_{\mathrm{nh}} = -i\gamma\sum_R \hat{S}^z_R.
\end{equation}

\noindent
and the Holstein-Primakoff transformation applied to our quantum system confined in the $z$ polarized phase gives:

\begin{align}
& \hat{S}^{z}_R = \frac{1}{2} - \hat{a}_R^{\dag} \hat{a}_R.
\end{align}

\noindent
This means that it is not possible to generate a non-Hermitian quartic term in terms of bosonic operators. As a consequence, even if we go beyond the LSWT by considering higher order terms in the Holstein-Primakoff transformation for $\hat{S}^x_R$ the $x$-axis spin operator on the lattice site $R$, the finite-time divergence can not be cured. 

\bibliographystyle{revtex}
\bibliography{biblioJD}

\begin{thebibliography}{10}
\providecommand*{\bibinfo}[2]{#2}
\providecommand*{\eprint}[1]{#1}
\providecommand*{\url}[1]{#1}
\bibitem{jurcevic2014}
\bibinfo{author}{P.~Jurcevic}, \bibinfo{author}{B.~P. Lanyon},
  \bibinfo{author}{P.~Hauke}, \bibinfo{author}{C.~Hempel},
  \bibinfo{author}{P.~Zoller}, \bibinfo{author}{R.~Blatt}, and
  \bibinfo{author}{C.~F. Roos}, \bibinfo{title}{\emph{Quasiparticle engineering
  and entanglement propagation in a quantum many-body system}},
  \bibinfo{journal}{Nature} \bibinfo{volume}{\textbf{511}},
  \bibinfo{pages}{202} (\bibinfo{date}{2014}).
\bibitem{richerme2014}
\bibinfo{author}{P.~Richerme}, \bibinfo{author}{Z.-X. Gong},
  \bibinfo{author}{A.~Lee}, \bibinfo{author}{C.~Senko},
  \bibinfo{author}{J.~Smith}, \bibinfo{author}{M.~Foss-Feig},
  \bibinfo{author}{S.~Michalakis}, \bibinfo{author}{A.~V. Gorshkov}, and
  \bibinfo{author}{C.~Monroe}, \bibinfo{title}{\emph{Non-local propagation of
  correlations in quantum systems with long-range interactions}},
  \bibinfo{journal}{Nature} \bibinfo{volume}{\textbf{511}},
  \bibinfo{pages}{198} (\bibinfo{date}{2014}).
\bibitem{cheneau2012}
\bibinfo{author}{M.~Cheneau}, \bibinfo{author}{P.~Barmettler},
  \bibinfo{author}{D.~Poletti}, \bibinfo{author}{M.~Endres},
  \bibinfo{author}{P.~Schauss}, \bibinfo{author}{T.~Fukuhara},
  \bibinfo{author}{C.~Gross}, \bibinfo{author}{I.~Bloch},
  \bibinfo{author}{C.~Kollath}, and \bibinfo{author}{S.~Kuhr},
  \bibinfo{title}{\emph{Light-cone-like spreading of correlations in a quantum
  many-body system}}, \bibinfo{journal}{\Jnature}
  \bibinfo{volume}{\textbf{481}}, \bibinfo{pages}{484} (\bibinfo{date}{2012}).
\bibitem{polkovnikov2011}
\bibinfo{author}{A.~Polkovnikov}, \bibinfo{author}{K.~Sengupta},
  \bibinfo{author}{A.~Silva}, and \bibinfo{author}{M.~Vengalattore},
  \bibinfo{title}{\emph{Colloquium: Nonequilibrium dynamics of closed
  interacting quantum systems}}, \bibinfo{journal}{Rev. Mod. Phys.}
  \bibinfo{volume}{\textbf{83}}, \bibinfo{pages}{863} (\bibinfo{date}{2011}).
\bibitem{gogolin2016}
\bibinfo{author}{C.~Gogolin} and \bibinfo{author}{J.~Eisert},
  \bibinfo{title}{\emph{Equilibration, thermalisation, and the emergence of
  statistical mechanics in closed quantum systems}}, \bibinfo{journal}{Reports
  on Progress in Physics} \bibinfo{volume}{\textbf{79}}(5),
  \bibinfo{pages}{056001} (\bibinfo{date}{2016}).
\bibitem{calabrese2005}
\bibinfo{author}{P.~Calabrese} and \bibinfo{author}{J.~Cardy},
  \bibinfo{title}{\emph{Evolution of entanglement entropy in one-dimensional
  systems}}, \bibinfo{journal}{Journal of Statistical Mechanics: Theory and
  Experiment} \bibinfo{volume}{\textbf{2005}}(04), \bibinfo{pages}{P04010}
  (\bibinfo{date}{2005}).
\bibitem{Eisler_2011}
\bibinfo{author}{V.~Eisler}, \bibinfo{title}{\emph{Crossover between ballistic
  and diffusive transport: the quantum exclusion process}},
  \bibinfo{journal}{Journal of Statistical Mechanics: Theory and Experiment}
  \bibinfo{volume}{\textbf{2011}}(06), \bibinfo{pages}{P06007}
  (\bibinfo{date}{jun 2011}).
\bibitem{kollath2018}
\bibinfo{author}{J.-S. Bernier}, \bibinfo{author}{R.~Tan},
  \bibinfo{author}{L.~Bonnes}, \bibinfo{author}{C.~Guo},
  \bibinfo{author}{D.~Poletti}, and \bibinfo{author}{C.~Kollath},
  \bibinfo{title}{\emph{Light-cone and diffusive propagation of correlations in
  a many-body dissipative system}}, \bibinfo{journal}{Physical Review Letters}
  \bibinfo{volume}{\textbf{120}}(2) (\bibinfo{date}{2018}).
\bibitem{turkeshi2021diffusion}
\bibinfo{author}{X.~Turkeshi} and \bibinfo{author}{M.~Schir\'o},
  \bibinfo{title}{\emph{Diffusion and thermalization in a boundary-driven
  dephasing model}}, \bibinfo{journal}{Phys. Rev. B}
  \bibinfo{volume}{\textbf{104}}, \bibinfo{pages}{144301} (\bibinfo{date}{Oct
  2021}).
\bibitem{alba2022}
\bibinfo{author}{V.~Alba} and \bibinfo{author}{F.~Carollo},
  \bibinfo{title}{\emph{Noninteracting fermionic systems with localized losses:
  Exact results in the hydrodynamic limit}}, \bibinfo{journal}{Physical Review
  B} \bibinfo{volume}{\textbf{105}}(5) (\bibinfo{date}{2022}).
\bibitem{rosso2021}
\bibinfo{author}{L.~Rosso}, \bibinfo{author}{D.~Rossini},
  \bibinfo{author}{A.~Biella}, and \bibinfo{author}{L.~Mazza},
  \bibinfo{title}{\emph{One-dimensional spin-1/2 fermionic gases with two-body
  losses: Weak dissipation and spin conservation}}, \bibinfo{journal}{Phys.
  Rev. A} \bibinfo{volume}{\textbf{104}}, \bibinfo{pages}{053305}
  (\bibinfo{date}{2021}).
\bibitem{rosso2023}
\bibinfo{author}{L.~Rosso}, \bibinfo{author}{A.~Biella},
  \bibinfo{author}{J.~De~Nardis}, and \bibinfo{author}{L.~Mazza},
  \bibinfo{title}{\emph{Dynamical theory for one-dimensional fermions with
  strong two-body losses: Universal non-hermitian zeno physics and spin-charge
  separation}}, \bibinfo{journal}{Phys. Rev. A} \bibinfo{volume}{\textbf{107}},
  \bibinfo{pages}{013303} (\bibinfo{date}{2023}).
\bibitem{mazza2023dissipative}
\bibinfo{author}{G.~Mazza} and \bibinfo{author}{M.~Schir\`o},
  \bibinfo{title}{\emph{Dissipative dynamics of a fermionic superfluid with
  two-body losses}}, \bibinfo{journal}{Phys. Rev. A}
  \bibinfo{volume}{\textbf{107}}, \bibinfo{pages}{L051301} (\bibinfo{date}{May
  2023}).
\bibitem{alba2020}
\bibinfo{author}{V.~Alba} and \bibinfo{author}{F.~Carollo},
  \bibinfo{title}{\emph{Spreading of correlations in markovian open quantum
  systems}}, \bibinfo{journal}{Physical Review B}
  \bibinfo{volume}{\textbf{103}}(2) (\bibinfo{date}{2021}).
\bibitem{breuer2007}
\bibinfo{author}{H.-P. Breuer} and \bibinfo{author}{F.~Petruccione},
  \bibinfo{title}{\emph{{The Theory of Open Quantum Systems}}}
  (\bibinfo{publisher}{Oxford University Press}, \bibinfo{year}{2007}).
\bibitem{ashida2020}
\bibinfo{author}{Y.~Ashida}, \bibinfo{author}{Z.~Gong}, and
  \bibinfo{author}{M.~Ueda}, \bibinfo{title}{\emph{Non-hermitian physics}},
  \bibinfo{journal}{Advances in Physics} \bibinfo{volume}{\textbf{69}}(3),
  \bibinfo{pages}{249} (\bibinfo{date}{2020}).
\bibitem{daley2014}
\bibinfo{author}{A.~J. Daley}, \bibinfo{title}{\emph{Quantum trajectories and
  open many-body quantum systems}}, \bibinfo{journal}{Advances in Physics}
  \bibinfo{volume}{\textbf{63}}(2), \bibinfo{pages}{77} (\bibinfo{date}{2014}).
\bibitem{Heiss_2012}
\bibinfo{author}{W.~D. Heiss}, \bibinfo{title}{\emph{The physics of exceptional
  points}}, \bibinfo{journal}{J. Phys. Math. Theor.}
  \bibinfo{volume}{\textbf{45}}, \bibinfo{pages}{444016}
  (\bibinfo{date}{2012}).
\bibitem{gong2018topological}
\bibinfo{author}{Z.~Gong}, \bibinfo{author}{Y.~Ashida},
  \bibinfo{author}{K.~Kawabata}, \bibinfo{author}{K.~Takasan},
  \bibinfo{author}{S.~Higashikawa}, and \bibinfo{author}{M.~Ueda},
  \bibinfo{title}{\emph{Topological phases of non-hermitian systems}},
  \bibinfo{journal}{Phys. Rev. X.} \bibinfo{volume}{\textbf{8}},
  \bibinfo{pages}{031079} (\bibinfo{date}{2018}).
\bibitem{bergholtz2021exceptional}
\bibinfo{author}{E.~J. Bergholtz}, \bibinfo{author}{J.~C. Budich}, and
  \bibinfo{author}{F.~K. Kunst}, \bibinfo{title}{\emph{Exceptional topology of
  non-hermitian systems}}, \bibinfo{journal}{Rev. Mod. Phys.}
  \bibinfo{volume}{\textbf{93}}, \bibinfo{pages}{015005}
  (\bibinfo{date}{2021}).
\bibitem{chen2023}
\bibinfo{author}{G.~Chen}, \bibinfo{author}{F.~Song}, and
  \bibinfo{author}{J.~L. Lado}, \bibinfo{title}{\emph{Topological spin
  excitations in non-hermitian spin chains with a generalized kernel polynomial
  algorithm}}, \bibinfo{journal}{Phys. Rev. Lett.}
  \bibinfo{volume}{\textbf{130}}, \bibinfo{pages}{100401}
  (\bibinfo{date}{2023}).
\bibitem{couvreur2017entanglement}
\bibinfo{author}{R.~Couvreur}, \bibinfo{author}{J.~L. Jacobsen}, and
  \bibinfo{author}{H.~Saleur}, \bibinfo{title}{\emph{Entanglement in nonunitary
  quantum critical spin chains}}, \bibinfo{journal}{Phys. Rev. Lett.}
  \bibinfo{volume}{\textbf{119}}, \bibinfo{pages}{040601}
  (\bibinfo{date}{2017}).
\bibitem{herviou2018entanglement}
\bibinfo{author}{L.~Herviou}, \bibinfo{author}{N.~Regnault}, and
  \bibinfo{author}{J.~H. Bardarson}, \bibinfo{title}{\emph{Entanglement
  spectrum and symmetries in non-hermitian fermionic non-interacting models}},
  \bibinfo{journal}{Scipost. Phys.} \bibinfo{volume}{\textbf{7}}
  (\bibinfo{date}{2019}).
\bibitem{chang2020entanglement}
\bibinfo{author}{P.-Y. Chang}, \bibinfo{author}{J.-S. You},
  \bibinfo{author}{X.~Wen}, and \bibinfo{author}{S.~Ryu},
  \bibinfo{title}{\emph{Entanglement spectrum and entropy in topological
  non-hermitian systems and nonunitary conformal field theory}},
  \bibinfo{journal}{Phys. Rev. Res.} \bibinfo{volume}{\textbf{2}},
  \bibinfo{pages}{033069} (\bibinfo{date}{2020}).
\bibitem{sarang2021}
\bibinfo{author}{S.~Gopalakrishnan} and \bibinfo{author}{M.~J. Gullans},
  \bibinfo{title}{\emph{Entanglement and purification transitions in
  non-hermitian quantum mechanics}}, \bibinfo{journal}{Phys. Rev. Lett.}
  \bibinfo{volume}{\textbf{126}}, \bibinfo{pages}{170503}
  (\bibinfo{date}{2021}).
\bibitem{turkeshi2023}
\bibinfo{author}{X.~Turkeshi} and \bibinfo{author}{M.~Schir{\'{o}}},
  \bibinfo{title}{\emph{Entanglement and correlation spreading in non-hermitian
  spin chains}}, \bibinfo{journal}{Physical Review B}
  \bibinfo{volume}{\textbf{107}}(2) (\bibinfo{date}{2023}).
\bibitem{legal2023}
\bibinfo{author}{Y.~L. Gal}, \bibinfo{author}{X.~Turkeshi}, and
  \bibinfo{author}{M.~Schir{\`{o}}}, \bibinfo{title}{\emph{Volume-to-area law
  entanglement transition in a non-hermitian free fermionic chain}},
  \bibinfo{journal}{{SciPost} Physics} \bibinfo{volume}{\textbf{14}}(5)
  (\bibinfo{date}{2023}).
\bibitem{kawabata2023entanglement}
\bibinfo{author}{K.~Kawabata}, \bibinfo{author}{T.~Numasawa}, and
  \bibinfo{author}{S.~Ryu}, \bibinfo{title}{\emph{Entanglement phase transition
  induced by the non-hermitian skin effect}}, \bibinfo{journal}{Phys. Rev. X}
  \bibinfo{volume}{\textbf{13}}, \bibinfo{pages}{021007}
  (\bibinfo{date}{2023}).
\bibitem{granet2023volume}
\bibinfo{author}{E.~Granet}, \bibinfo{author}{C.~Zhang}, and
  \bibinfo{author}{H.~Dreyer}, \bibinfo{title}{\emph{Volume-law to area-law
  entanglement transition in a nonunitary periodic gaussian circuit}},
  \bibinfo{journal}{Phys. Rev. Lett.} \bibinfo{volume}{\textbf{130}},
  \bibinfo{pages}{230401} (\bibinfo{date}{Jun 2023}).
\bibitem{su2023dynamics}
\bibinfo{author}{L.~Su}, \bibinfo{author}{A.~Clerk}, and
  \bibinfo{author}{I.~Martin}, \bibinfo{title}{\emph{Dynamics and phases of
  nonunitary floquet transverse-field ising model}} (\bibinfo{date}{2023}),
  \eprint{2306.07428}.
\bibitem{zhang2023antiunitary}
\bibinfo{author}{C.~Zhang} and \bibinfo{author}{E.~Granet},
  \bibinfo{title}{\emph{Antiunitary symmetry breaking and a hierarchy of
  purification transitions in floquet non-unitary circuits}}
  (\bibinfo{date}{2023}), \eprint{2307.07003}.
\bibitem{dora2020quantum}
\bibinfo{author}{B.~D\'ora} and \bibinfo{author}{C.~P. Moca},
  \bibinfo{title}{\emph{Quantum quench in $\mathcal{P}\mathcal{T}$-symmetric
  luttinger liquid}}, \bibinfo{journal}{Phys. Rev. Lett.}
  \bibinfo{volume}{\textbf{124}}, \bibinfo{pages}{136802}
  (\bibinfo{date}{2020}).
\bibitem{banerjee2023emergent}
\bibinfo{author}{T.~Banerjee} and \bibinfo{author}{K.~Sengupta},
  \bibinfo{title}{\emph{Emergent conservation in the floquet dynamics of
  integrable non-hermitian models}}, \bibinfo{journal}{Phys. Rev. B}
  \bibinfo{volume}{\textbf{107}}, \bibinfo{pages}{155117} (\bibinfo{date}{Apr
  2023}).
\bibitem{hatano1996localization}
\bibinfo{author}{N.~Hatano} and \bibinfo{author}{D.~R. Nelson},
  \bibinfo{title}{\emph{Localization transitions in non-hermitian quantum
  mechanics}}, \bibinfo{journal}{Phys. Rev. Lett.}
  \bibinfo{volume}{\textbf{77}}, \bibinfo{pages}{570} (\bibinfo{date}{1996}).
\bibitem{zhang2022symmetry}
\bibinfo{author}{S.-B. Zhang}, \bibinfo{author}{M.~M. Denner},
  \bibinfo{author}{T.~c.~v. Bzdu\ifmmode~\check{s}\else \v{s}\fi{}ek},
  \bibinfo{author}{M.~A. Sentef}, and \bibinfo{author}{T.~Neupert},
  \bibinfo{title}{\emph{Symmetry breaking and spectral structure of the
  interacting hatano-nelson model}}, \bibinfo{journal}{Phys. Rev. B}
  \bibinfo{volume}{\textbf{106}}, \bibinfo{pages}{L121102}
  (\bibinfo{date}{2022}).
\bibitem{starkov2023quantum}
\bibinfo{author}{G.~A. Starkov}, \bibinfo{author}{M.~V. Fistul}, and
  \bibinfo{author}{I.~M. Eremin}, \bibinfo{title}{\emph{Quantum phase
  transitions in non-hermitian pt-symmetric transverse-field ising spin
  chains}}, \bibinfo{journal}{Annals of Physics} \bibinfo{pages}{p. 169268}
  (\bibinfo{date}{2023}).
\bibitem{ghosh2023hilbert}
\bibinfo{author}{S.~Ghosh}, \bibinfo{author}{K.~Sengupta}, and
  \bibinfo{author}{I.~Paul}, \bibinfo{title}{\emph{Hilbert space fragmentation
  imposed real spectrum of a non-hermitian system}} (\bibinfo{date}{2023}),
  \eprint{2307.05679}.
\bibitem{yamamoto2022universal}
\bibinfo{author}{K.~Yamamoto}, \bibinfo{author}{M.~Nakagawa},
  \bibinfo{author}{M.~Tezuka}, \bibinfo{author}{M.~Ueda}, and
  \bibinfo{author}{N.~Kawakami}, \bibinfo{title}{\emph{Universal properties of
  dissipative tomonaga-luttinger liquids: Case study of a non-hermitian xxz
  spin chain}}, \bibinfo{journal}{Phys. Rev. B} \bibinfo{volume}{\textbf{105}},
  \bibinfo{pages}{205125} (\bibinfo{date}{2022}).
\bibitem{kattel2023exact}
\bibinfo{author}{P.~Kattel}, \bibinfo{author}{P.~R. Pasnoori}, and
  \bibinfo{author}{N.~Andrei}, \bibinfo{title}{\emph{Exact solution of a
  non-hermitian -symmetric spin chain}}, \bibinfo{journal}{Journal of Physics
  A: Mathematical and Theoretical} \bibinfo{volume}{\textbf{56}}(32),
  \bibinfo{pages}{325001} (\bibinfo{date}{2023}).
\bibitem{durr2009}
\bibinfo{author}{S.~D\"urr}, \bibinfo{author}{J.~J. Garc\'{\i}a-Ripoll},
  \bibinfo{author}{N.~Syassen}, \bibinfo{author}{D.~M. Bauer},
  \bibinfo{author}{M.~Lettner}, \bibinfo{author}{J.~I. Cirac}, and
  \bibinfo{author}{G.~Rempe}, \bibinfo{title}{\emph{Lieb-liniger model of a
  dissipation-induced tonks-girardeau gas}}, \bibinfo{journal}{Phys. Rev. A}
  \bibinfo{volume}{\textbf{79}}, \bibinfo{pages}{023614}
  (\bibinfo{date}{2009}).
\bibitem{holstein1940field}
\bibinfo{author}{T.~Holstein} and \bibinfo{author}{H.~Primakoff},
  \bibinfo{title}{\emph{Field dependence of the intrinsic domain magnetization
  of a ferromagnet}}, \bibinfo{journal}{Phys. Rev.}
  \bibinfo{volume}{\textbf{58}}, \bibinfo{pages}{1098} (\bibinfo{date}{1940}).
\bibitem{dyson1956general}
\bibinfo{author}{F.~J. Dyson}, \bibinfo{title}{\emph{General theory of
  spin-wave interactions}}, \bibinfo{journal}{Phys. Rev.}
  \bibinfo{volume}{\textbf{102}}, \bibinfo{pages}{1217} (\bibinfo{date}{1956}).
\bibitem{sandri2012linear}
\bibinfo{author}{M.~Sandri}, \bibinfo{author}{M.~Schir\'o}, and
  \bibinfo{author}{M.~Fabrizio}, \bibinfo{title}{\emph{Linear ramps of
  interaction in the fermionic hubbard model}}, \bibinfo{journal}{Phys. Rev. B}
  \bibinfo{volume}{\textbf{86}}, \bibinfo{pages}{075122} (\bibinfo{date}{Aug
  2012}).
\bibitem{lerose2019impact}
\bibinfo{author}{A.~Lerose}, \bibinfo{author}{B.~\ifmmode \check{Z}\else
  \v{Z}\fi{}unkovi\ifmmode~\check{c}\else \v{c}\fi{}},
  \bibinfo{author}{J.~Marino}, \bibinfo{author}{A.~Gambassi}, and
  \bibinfo{author}{A.~Silva}, \bibinfo{title}{\emph{Impact of nonequilibrium
  fluctuations on prethermal dynamical phase transitions in long-range
  interacting spin chains}}, \bibinfo{journal}{Phys. Rev. B}
  \bibinfo{volume}{\textbf{99}}, \bibinfo{pages}{045128} (\bibinfo{date}{Jan
  2019}).
\bibitem{despres2018}
\bibinfo{author}{L.~Cevolani}, \bibinfo{author}{J.~Despres},
  \bibinfo{author}{G.~Carleo}, \bibinfo{author}{L.~Tagliacozzo}, and
  \bibinfo{author}{L.~Sanchez-Palencia}, \bibinfo{title}{\emph{Universal
  scaling laws for correlation spreading in quantum systems with short- and
  long-range interactions}}, \bibinfo{journal}{Phys. Rev. B}
  \bibinfo{volume}{\textbf{98}}, \bibinfo{pages}{024302}
  (\bibinfo{date}{2018}).
\bibitem{despres2019}
\bibinfo{author}{J.~Despres}, \bibinfo{author}{L.~Villa}, and
  \bibinfo{author}{L.~Sanchez-Palencia}, \bibinfo{title}{\emph{Twofold
  correlation spreading in a strongly correlated lattice bose gas}},
  \bibinfo{journal}{Scientific Reports} \bibinfo{volume}{\textbf{9}}(1)
  (\bibinfo{date}{2019}).
\bibitem{despres2021}
\bibinfo{author}{J.~T. Schneider}, \bibinfo{author}{J.~Despres},
  \bibinfo{author}{S.~J. Thomson}, \bibinfo{author}{L.~Tagliacozzo}, and
  \bibinfo{author}{L.~Sanchez-Palencia}, \bibinfo{title}{\emph{Spreading of
  correlations and entanglement in the long-range transverse ising chain}},
  \bibinfo{journal}{Physical Review Research} \bibinfo{volume}{\textbf{3}}(1)
  (\bibinfo{date}{2021}).
\bibitem{barmettler2012}
\bibinfo{author}{P.~Barmettler}, \bibinfo{author}{D.~Poletti},
  \bibinfo{author}{M.~Cheneau}, and \bibinfo{author}{C.~Kollath},
  \bibinfo{title}{\emph{Propagation front of correlations in an interacting
  bose gas}}, \bibinfo{journal}{Physical Review A}
  \bibinfo{volume}{\textbf{85}}(5) (\bibinfo{date}{2012}).
\bibitem{hauke2013}
\bibinfo{author}{P.~Hauke} and \bibinfo{author}{L.~Tagliacozzo},
  \bibinfo{title}{\emph{Spread of correlations in long-range interacting
  quantum systems}}, \bibinfo{journal}{Physical Review Letters}
  \bibinfo{volume}{\textbf{111}}(20) (\bibinfo{date}{2013}).
\bibitem{natu2013}
\bibinfo{author}{S.~S. Natu} and \bibinfo{author}{E.~J. Mueller},
  \bibinfo{title}{\emph{Dynamics of correlations in a dilute bose gas following
  an interaction quench}}, \bibinfo{journal}{Physical Review A}
  \bibinfo{volume}{\textbf{87}}(5) (\bibinfo{date}{2013}).
\bibitem{cevolani2015}
\bibinfo{author}{L.~Cevolani}, \bibinfo{author}{G.~Carleo}, and
  \bibinfo{author}{L.~Sanchez-Palencia}, \bibinfo{title}{\emph{Protected
  quasilocality in quantum systems with long-range interactions}},
  \bibinfo{journal}{Physical Review A} \bibinfo{volume}{\textbf{92}}(4)
  (\bibinfo{date}{2015}).
\bibitem{cevolani2016}
\bibinfo{author}{L.~Cevolani}, \bibinfo{author}{G.~Carleo}, and
  \bibinfo{author}{L.~Sanchez-Palencia}, \bibinfo{title}{\emph{Spreading of
  correlations in exactly solvable quantum models with long-range interactions
  in arbitrary dimensions}}, \bibinfo{journal}{New Journal of Physics}
  \bibinfo{volume}{\textbf{18}}(9), \bibinfo{pages}{093002}
  (\bibinfo{date}{2016}).
\bibitem{buyskikh2016}
\bibinfo{author}{A.~S. Buyskikh}, \bibinfo{author}{M.~Fagotti},
  \bibinfo{author}{J.~Schachenmayer}, \bibinfo{author}{F.~Essler}, and
  \bibinfo{author}{A.~J. Daley}, \bibinfo{title}{\emph{Entanglement growth and
  correlation spreading with variable-range interactions in spin and fermionic
  tunneling models}}, \bibinfo{journal}{Physical Review A}
  \bibinfo{volume}{\textbf{93}}(5) (\bibinfo{date}{2016}).
\bibitem{frerot2018}
\bibinfo{author}{I.~Fr{\'{e}}rot}, \bibinfo{author}{P.~Naldesi}, and
  \bibinfo{author}{T.~Roscilde}, \bibinfo{title}{\emph{Multispeed
  prethermalization in quantum spin models with power-law decaying
  interactions}}, \bibinfo{journal}{Physical Review Letters}
  \bibinfo{volume}{\textbf{120}}(5) (\bibinfo{date}{2018}).
\bibitem{meden2023}
\bibinfo{author}{V.~Meden}, \bibinfo{author}{L.~Grunwald}, and
  \bibinfo{author}{D.~M. Kennes}, \bibinfo{title}{\emph{Pt-symmetric,
  non-hermitian quantum many-body physics - a methodological perspective}},
  \bibinfo{journal}{Reports on Progress in Physics}
  \bibinfo{volume}{\textbf{86}}(12) (\bibinfo{date}{2023}).
\bibitem{ashida2018}
\bibinfo{author}{Y.~Ashida} and \bibinfo{author}{M.~Ueda},
  \bibinfo{title}{\emph{Full-counting many-particle dynamics: Nonlocal and
  chiral propagation of correlations}}, \bibinfo{journal}{Phys. Rev. Lett.}
  \bibinfo{volume}{\textbf{120}}, \bibinfo{pages}{185301}
  (\bibinfo{date}{2018}).
\bibitem{moca2021}
\bibinfo{author}{C.~P. Moca} and \bibinfo{author}{B.~D\'ora},
  \bibinfo{title}{\emph{Universal conductance of a $\mathcal{PT}$-symmetric
  luttinger liquid after a quantum quench}}, \bibinfo{journal}{Phys. Rev. B}
  \bibinfo{volume}{\textbf{104}}, \bibinfo{pages}{125124}
  (\bibinfo{date}{2021}).
\bibitem{lee2014bis}
\bibinfo{author}{T.~E. Lee} and \bibinfo{author}{C.-K. Chan},
  \bibinfo{title}{\emph{Heralded magnetism in non-hermitian atomic systems}},
  \bibinfo{journal}{Phys. Rev. X} \bibinfo{volume}{\textbf{4}},
  \bibinfo{pages}{041001} (\bibinfo{date}{2014}).
\bibitem{lee2014}
\bibinfo{author}{T.~E. Lee}, \bibinfo{author}{F.~Reiter}, and
  \bibinfo{author}{N.~Moiseyev}, \bibinfo{title}{\emph{Entanglement and spin
  squeezing in non-hermitian phase transitions}}, \bibinfo{journal}{Physical
  Review Letters} \bibinfo{volume}{\textbf{113}}(25) (\bibinfo{date}{2014}).
\bibitem{hickey2013}
\bibinfo{author}{J.~M. Hickey}, \bibinfo{author}{S.~Genway},
  \bibinfo{author}{I.~Lesanovsky}, and \bibinfo{author}{J.~P. Garrahan},
  \bibinfo{title}{\emph{Time-integrated observables as order parameters for
  full counting statistics transitions in closed quantum systems}},
  \bibinfo{journal}{Phys. Rev. B} \bibinfo{volume}{\textbf{87}},
  \bibinfo{pages}{184303} (\bibinfo{date}{2013}).
\bibitem{zhang2012}
\bibinfo{author}{X.~Z. Zhang} and \bibinfo{author}{Z.~Song},
  \bibinfo{title}{\emph{Non-hermitian anisotropic $xy$ model with intrinsic
  rotation-time-reversal symmetry}}, \bibinfo{journal}{Phys. Rev. A}
  \bibinfo{volume}{\textbf{87}}, \bibinfo{pages}{012114}
  (\bibinfo{date}{2013}).
\bibitem{lieb1972}
\bibinfo{author}{E.~H. Lieb} and \bibinfo{author}{D.~W. Robinson},
  \bibinfo{title}{\emph{The finite group velocity of quantum spin systems}},
  \bibinfo{journal}{\JCommMathPhys} \bibinfo{volume}{\textbf{28}},
  \bibinfo{pages}{251} (\bibinfo{date}{1972}).

\end{thebibliography}

\end{document}